\documentclass{aastex631}
\usepackage{graphicx}
\usepackage{amsmath}
\usepackage{xcolor}
\usepackage{bm}

\begin{document}

\title{Adaptive Detection of Fast Moving Celestial Objects Using a Mixture of Experts and Physical-Inspired Neural Network}

\correspondingauthor{Peng Jia}
\email{robinmartin20@gmail.com}

\author[0000-0001-6623-0931]{Peng Jia}
\affiliation{College of Physics and Optoelectronics, Taiyuan University of Technology, Taiyuan, 030024, China}

\author{Ge Li}
\affiliation{College of Physics and Optoelectronics, Taiyuan University of Technology, Taiyuan, 030024, China}

\author{Bafeng Cheng}
\affiliation{College of Physics and Optoelectronics, Taiyuan University of Technology, Taiyuan, 030024, China}

\author{Yushan Li}
\affiliation{College of Physics and Optoelectronics, Taiyuan University of Technology, Taiyuan, 030024, China}

\author{Rongyu Sun}
\affiliation{Key Laboratory of Space Object and Debris Observation, Purple Mountain Observatory, Chinese Academy of Sciences, Nanjing, 210023, China}



\begin{abstract}
Fast moving celestial objects are characterized by velocities across the celestial sphere that significantly differ from the motions of background stars. In observational images, these objects exhibit distinct shapes, contrasting with the typical appearances of stars. Depending on the observational method employed, these celestial entities may be designated as near-Earth objects or asteroids. Historically, fast moving celestial objects have been observed using ground-based telescopes, where the relative stability of stars and Earth facilitated effective image differencing techniques alongside traditional fast moving celestial object detection and classification algorithms. However, the growing prevalence of space-based telescopes, along with their diverse observational modes, produces images with different properties, rendering conventional methods less effective. This paper presents a novel algorithm for detecting fast moving celestial objects within star fields. Our approach enhances state-of-the-art fast moving celestial object detection neural networks by transforming them into physical-inspired neural networks. These neural networks leverage the point spread function of the telescope and the specific observational mode as prior information; they can directly identify moving fast moving celestial objects within star fields without requiring additional training, thereby addressing the limitations of traditional techniques. Additionally, all neural networks are integrated using the mixture of experts technique, forming a comprehensive fast moving celestial object detection algorithm. We have evaluated our algorithm using simulated observational data that mimics various observations carried out by space based telescope scenarios and real observation images. Results demonstrate that our method effectively detects fast moving celestial objects across different observational modes and telescope configurations.
\end{abstract}


\section{Introduction} \label{sec:intro}
All celestial objects in the universe are in motion. Due to the vast distances separating Earth from stars, their positions in images captured by ground based telescopes appear relatively stable during observations. When a telescope is set to a fixed location in the celestial sphere, known as sidereal mode, the positional changes of these celestial bodies can only be detected by comparing images taken over several years. These changes arise from Earth’s precession and nutation, as well as the proper motion of the stars \citep{woolard2012spherical}. Frequently, these stars are referred to as background star fields during observations. In contrast, celestial objects within our solar system exhibit significantly faster motion. For instance, planets such as Mars, Jupiter, and Venus can move several arcmin per day, which may be discerned using high resolution telescopes \citep{kovalevsky1991scientific}. Objects that are closer to Earth or that possess greater speeds will traverse the image plane even faster. Comets and asteroids, when in close proximity to Earth, can move at rates of several arcsec per second \citep{lowry2003ccd, kruzins2023deep}. Space debris and meteors represent the fastest moving celestial objects when observed from the earth, potentially traveling several arcmin per second \citep{sun2022precise, dignam2023space}.\\

Fast moving celestial objects (FMCB) possess unique scientific value. Firstly, tracking positions and orbits of space debris provides astronomers with crucial data on their movements, velocities, and trajectories. This information is instrumental in constructing accurate models of celestial mechanics and understanding gravitational interactions between various celestial bodies \citep{kazemi2024orbit}. Secondly, the detection of near-Earth objects (NEOs), particularly potentially hazardous asteroids and comets, enables scientists to assess impact risks and develop mitigation strategies \citep{Aaron2021, Daly_2023, mainzer2011preliminary}. This aspect of celestial observation has significant implications for Earth's long-term safety and the future of human civilization. Thirdly, the study of comets and meteors may offer insights into the potential for extraterrestrial life, because they carry materials such as water and amino acid in our solar system \citep{altwegg2016prebiotic, hanni2023oxygen}. Depending on the observation mode, exposure time, field of view, and pixel scale of telescopes, FMCB may encompass various celestial objects. When celestial objects are observed with longer exposure time and a smaller pixel scale, FMCB are often asteroids. Conversely, when the observation is conducted with shorter exposure time and a larger pixel scale, FMCB typically refer to space debris and meteors. In this paper, we do not classify FMCB based on their specific scientific objectives. Instead, we categorize FMCB that exhibit shapes significantly different from those of stars in observational images. Detection of these FMCB presents significant challenges. These FMCB are often extremely dim, as they do not emit light themselves and are only visible through reflected photons from their surfaces. Additionally, they can approach from any direction in space, further complicating observation efforts.\\

To tackle these challenges effectively, we urgently need a highly efficient detection framework. This system should seamlessly integrate both remote observational instruments and close range investigation technologies. Such an approach could involve a combination of ground based and space based telescopes, along with space shuttles for nearby observation and potentially even sample collection \citep{schmidt2018planetary}. In particular, space based telescopes are important, because they might be mounted on satellites or aboard space shuttles capable of traveling to and approaching these objects. As observational instruments advance, so do data processing techniques, with a robust data processing framework already in practice. Traditionally, astronomers employ image differencing and FMCB detection algorithms to identify moving FMCB candidates \citep{mehrholz2002detecting}. In practical applications, SExtractor identifies groups of pixels whose gray values surpass this threshold, with the additional criterion that the number of such pixels must also exceed a predetermined minimum \citep{bertin1996sextractor}. Once astronomers identify potential candidates of FMCB, they typically engage in manual vetting of the detection results, employ strategies that use confirmation across multiple frames \citep{ye2019toward, doubek2019neostel} or segment images of FMCB with streak spread function \citep{sharma2023astreaks}. These methods allow for the direct identification of such objects from observational images. Nonetheless, due to the substantial increase in the number and the mode of observations of FMCB in recent years, manually vetting these results has become impractical, verifying detection results through multiple observations has turned increasingly costly and building streak spread function is hard for images of FMCB with low signal to noise ratio.\\

Advancements in machine learning in recent years have significantly expanded its application in detecting FMCB. \citet{fraser2025detecting} have curated a collection of studies that highlight recent progress in identifying small bodies within the solar system using machine learning techniques. As discussed above, the challenge of detecting small bodies within the solar system—a specific subset of FMCB—is analogous to the broader problem of their identification in technological contexts. This similarity underscores the general framework applicable to both cases. Initially, the detection framework has incorporated candidate identification through image differencing, traditional source detection techniques, and machine learning based classification algorithms. Notably, the classification algorithms for FMCB have been directly adapted from state-of-the-art image classification methods \citep{waszczak2017small, keys2019digest2, jia2019optical,turpin2020vetting, panes2021identification, yan2022weak}. \citet{turpin2020vetting} have introduced a standardized framework for searching for optical transients in Ground-based Wide Angle Camera (GWAC) data, which has been subsequently enhanced by applying convolutional neural network methods for classification. \citet{duev2019deepstreaks} have introduced DeepStreaks, a deep learning system based on several convolutional neural networks (CNNs), including VGG6, resnet50, densenet121, specifically designed to identify rapidly NEOs in the Zwicky Transient Facility data. The conventional framework—utilizing traditional techniques to detect candidates of FMCB followed by machine learning algorithms for candidate classification—suffers from significant limitations. Specifically, the image differencing method combined with traditional target detection algorithms tends to create discontinuities in images of faint targets. This issue not only results in a failure to detect these targets but also leads to a high incidence of false detections.\\

\begin{figure*}
 \centering
 \begin{minipage}{0.25\linewidth}
    \includegraphics[width=\linewidth]{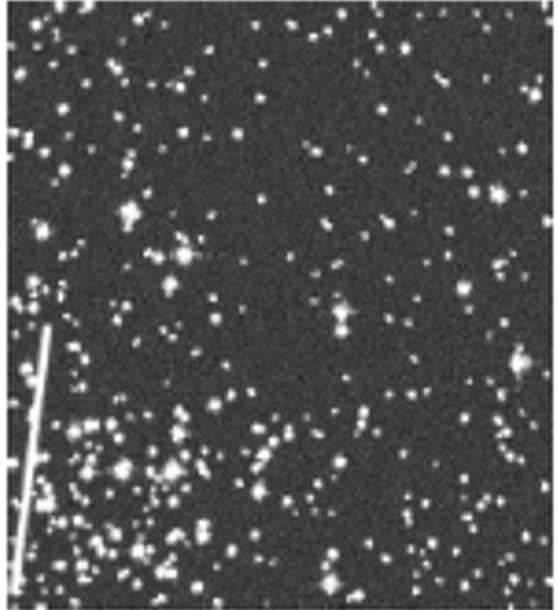}
    \centering\footnotesize\textbf{(a)} 
  \end{minipage}%
  \hspace{0.03\linewidth}%
  \begin{minipage}{0.25\linewidth}
    \includegraphics[width=\linewidth]{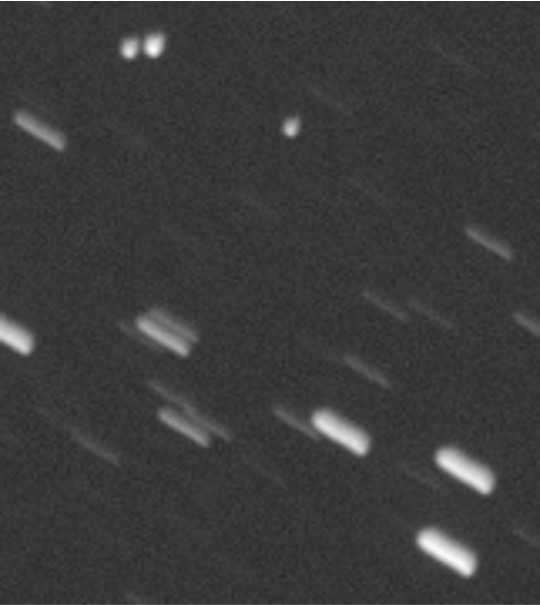}
    \centering\footnotesize\textbf{(b)}
  \end{minipage}%
  \hspace{0.03\linewidth}%
  \begin{minipage}{0.25\linewidth}
    \includegraphics[width=\linewidth]{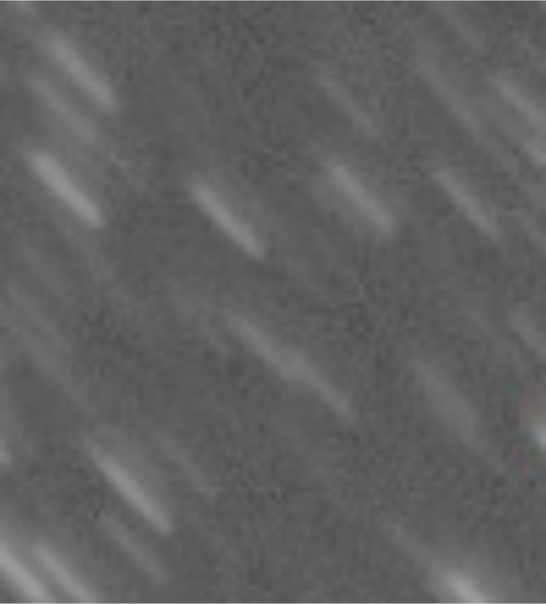}
    \centering\footnotesize\textbf{(c)} 
  \end{minipage}
  \caption{Three simulated images represent three different observation modes. These image are simulated using Skymaker \citep{bertin2009skymaker}. The panels, from left to right, illustrate: (a) Sidereal tracking mode: This image is generated with a simulated telescope that has a 1-meter diameter and a 1-degree field of view, operating in sidereal tracking mode with an exposure time of 0.5 seconds. (b) Object tracking mode: This image is created using a simulated telescope with a 60 cm diameter and a 0.6-degree field of view, set to object tracking mode with an exposure time of 0.4 seconds. (c) Intersection mode: This image is produced with a simulated telescope with a 30 cm diameter and a 0.6-degree field of view, employing intersection mode with an exposure time of 0.1 seconds.}
  \label{Figureinstruction}
\end{figure*}

End-to-end target detection neural networks include a single neural network to directly detect celestial objects from observation images, which seamlessly integrate the source detection and classification procedure. The end-to-end target detection neural networks are similar to human vision system, which is robust to images with low signal to noise ratio.  For example, \citet{lieu2019detecting} have applied neural networks to detect solar system objects, which uses transfer learning and retrained three CNN architectures, namely Inception v4, MobileNet\_v1\_1.0-224, and NASNet-A\_Large\_331, to classify asteroids and objects such as stars, galaxies, and cosmic rays. \citet{kruk2022hubble} used a citizen science project and deep learning techniques to search for trails of  Small Solar System Objects (SSOs)in Hubble Space Telescope (HST) data. \citet{cowan2023towards} have proposed a model consisting of an ensemble of five classifiers based on convolutional neural networks to classify asteroids, and have used the YOLOv4 \citep{bochkovskiy2020yolov4} object detection framework to locate asteroid trails. \citet{jeffries2024detection} have proposed a deep learning model based on the U-Net \citep{ronneberger2015u} architecture, an image segmentation technique, to detect and mask satellite streaks in astronomical data. Nonetheless, the aforementioned machine learning algorithms, which primarily focus on the morphology of FMCB images, are constrained by the available training datasets. When observational conditions change, these neural networks necessitate fine-tuning or even complete re-training. For ground-based observations, the telescope's point spread function (PSF) is influenced by atmospheric turbulence, resulting in variations between observations. For both space-based and ground-based observations, alterations in the observation mode, as demonstrated in Figure~\ref{Figureinstruction}, often require the collection of new training data and the re-training of the neural network. These challenges hinder the deployment of various algorithms in projects aimed at detecting FMCB.\\

To address the aforementioned challenges, our study  aims to address these limitations by combining the strengths of machine learning with physical-inspired models, ensuring robust performance across both simulated and real data. We propose an adaptive algorithm for detecting FMCB that incorporates more prior information, besides the morphology features, into a neural network framework. This innovative approach integrates key observational parameters, including the PSF - which represents the imaging capabilities of the optical telescope - and the relative motion between the telescope and star fields, directly into the neural network architecture. This design transforms the source detection algorithm into a physical-inspired, multi-modal neural network that can adapt to changing observation conditions or modes without requiring additional training. To further enhance the detection capabilities of our algorithm, we have implemented a mixture of experts framework (MoE) within the neural network, incorporating several state-of-the-art deep neural networks for source detection. This comprehensive design enables our neural network to efficiently detect FMCB against complex star field backgrounds. The structure of this paper is as follows: Section 2 presents a detailed description of our neural network architecture. Section 3 presents a statistical evaluation of our method's performance, illustrates its application across different simulated observational scenarios, and analyzes its effectiveness in processing real observation data. In Section 4, we conclude by summarizing our findings and suggesting potential directions for future research.\\

\section{Main design of the Algorithm} \label{sec2}
Our algorithm incorporates two innovative designs: a physical-inspired neural network PiNN architecture for FMCB detection and an overall detection framework based on the MoE approach. The PiNN proposed in this paper introduces two novel features. Firstly, it integrates the PSF of the telescope and the relative motion between the pointing direction of the telescope and star fields as prior information. Secondly, rather than concentrating on detecting labeled stars within observational images, this network is engineered to identify celestial objects that exhibit motion patterns differing from those of the labeled stars. This capability allows the network to detect FMCB with behaviors that are not represented within the training set. Regarding the MoE, we have redesigned several state-of-the-art FMCB detection neural networks, adapting them to our physical-inspired paradigm. These modified networks are then integrated into a comprehensive neural network system. With the design of the MoE, neural networks with different capabilities and designs can complement each other in terms of performance. The following subsections will delve into the specifics of these components, providing a detailed exploration of their design and functionality.\\

\subsection{Design Concept of the PiNN} \label{sec2.1}
Contemporary neural networks designed for FMCB detection primarily adhere to the supervised learning paradigm, as discussed in Section~\ref{sec:intro}. In principle, given a sufficiently complex neural network and a comprehensive dataset of images with corresponding labels, one could train the neural network to detect celestial objects from observational images. This approach has been extensively explored for celestial objects detection. However, when applied to detection of FMCB, this paradigm faces significant challenges. Firstly, the PSF and observation strategies often vary between observations, making it impractical for the training data to encompass all possible states. Moreover, FMCB exhibit diverse velocities and trajectories, rendering it unfeasible to train a neural network on a limited set of circumstances and expect it to generalize to objects with arbitrary motion characteristics. In fact, these issues have already become prominent even in the more established fields of astrometry and photometry of celestial objects \citep{sun2023pnet}. The fundamental principle of object detection neural networks is to generate multiple candidate FMCB and subsequently ascertain detection results based on the features of labeled fast-moving celestial objects. By exploring the intricacies of detecting these objects, we acknowledge that, while we lack knowledge of the appearance of FMCB across various observational images, we are familiar with the shapes of stars, which are not our focus of interest. Utilizing this knowledge, we can improve the neural network's capability to detect FMCB through the following design enhancements. \\

By revisiting the physical process involved in detecting FMCB, we can gain valuable insights for new designs. In the past, the source detection algorithm aimed to identify all celestial objects by analyzing images with pixels that have gray scale values exceeding a specified threshold, followed by classification through a trained neural network. However, this approach does not take into account all the physical information that could be utilized during the processing of observational images. For instance, the PSF of an optical system characterizes the impulse response of optical telescopes and can be determined prior to conducting observations \citep{racine1996telescope, wang2018automated, zhan2022database}. Additionally, the relative shift between the telescope's pointing direction and the star fields can be calculated after the observations. Together, the PSF and this relative shift provide a comprehensive description of how a star should appear in the observational images. If we can design an innovative neural network capable of detecting celestial objects while filtering out detection results resembling stars, which could be characterized as the convolution results between the PSF and a specified motion vector, we can distinctly identify all moving FMCB. This approach could effectively create a neural network optimized for the efficient detection of FMCB.\\

\begin{figure}
    \centering
    \includegraphics[scale=0.4]{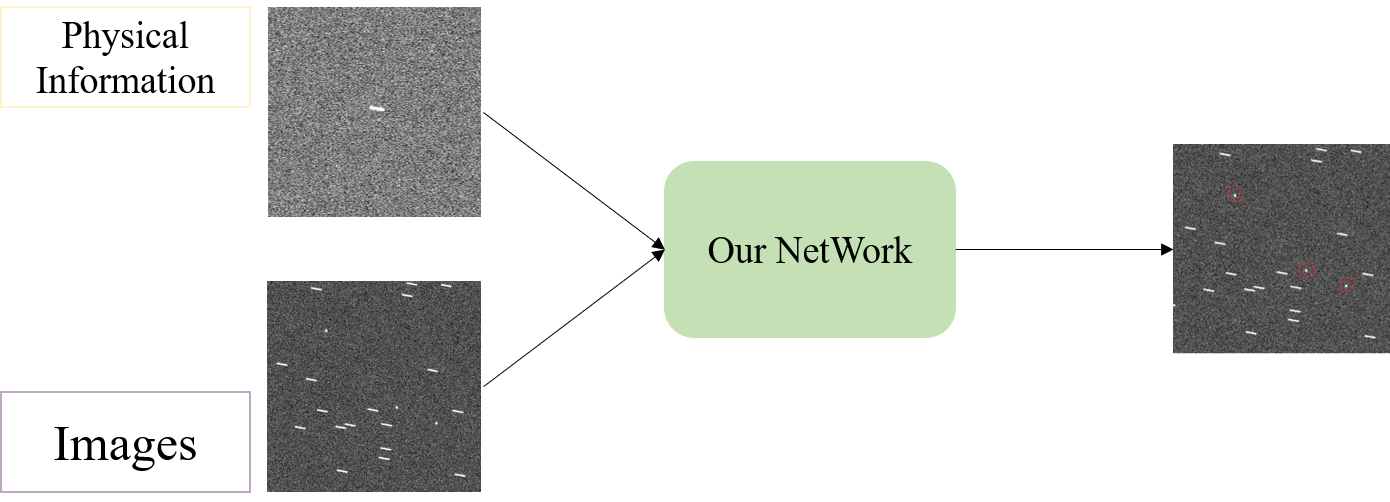}
    \caption{The schematic draw of the structure of the PiNN proposed in this paper.}
    \label{Figure1}
    \end{figure}

Building upon the aforementioned concept, we propose integrating physical information into the neural network architecture. Figure~\ref{Figure1} illustrates the structure of the PiNN. In contrast to conventional object detection neural networks, our model is designed to process not only image data but also physical parameters.  By incorporating both images and physical informed parameters, the neural network achieves a more comprehensive understanding of the celestial object detection problem, potentially enhancing its capacity to accurately identify and characterize FMCB. As depicted in Figure~\ref{Figure1}, the PSF and the vector to represent the observation mode are encoded into the image using Equation~\ref{eq1},
\begin{equation}
\bm{Encoded~Img = Img \ast PSF}
    \label{eq1}
\end{equation}
The nature of $Img$ varies depending on the telescope's operational mode. When the telescope operates in sidereal mode, $Img$ manifests as a point-like image. When the telescope functions in tracking mode or intersection mode, $Img$ takes the form of a streak. The characteristics of this streak are determined by three key factors: the exposure time, the relative velocity of the telescope's motion compared to the star fields, and the image's pixel scale. These relationships are mathematically expressed in Equation~\ref{eq2}, where $Dx$ and $Dy$ stand for pixel number along x and y direction, $v_x$ and $v_y$ stand for the relative velocity of the telescope's motion compared to the star background in the x and y directions, $t_{exp}$ stands for the length of exposure time and $pixelscale$ stands for the pixelscale of the obervation images. The PSF in Equation~\ref{eq1} does not include the transforms mentioned in Equation~\ref{eq2}, and the transformations in Equation~\ref{eq2} are applied to $Img$.
\begin{equation}
\begin{split}
Dx = v_x \times t_{exp} / pixelscale, \\
Dy = v_y \times t_{exp} / pixelscale .
\end{split}
\label{eq2}
\end{equation}

\subsection{Design Concept of the MoE for Fast Moving Celestial Object Detection} \label{sec2.2}
As FMCB detection neural networks grow increasingly complex and incorporate more parameters, their capabilities continue to increase. However, the question of which neural network architecture performs optimally in detecting FMCB remains open. This challenge is particularly pronounced when considering FMCB with diverse characteristics, such as those appearing as small point-like FMCB or moderate-sized streak-like FMCB. There is a significant risk that adapting a source detection neural network with a specific structure may lead to diminished performance across this varied range of object types. A potential solution for processing images with such diverse properties is to employ multiple neural networks for FMCB detection. This concept, known as the MoE, is a deep learning architecture designed to handle complex tasks \citep{masoudnia2014mixture, zhai2023smartMoE, chowdhury2023patch}. The MoE model comprises multiple expert models, each an independent neural network structure that may share similar or different architectures \citep{yuksel2012twenty}. In an MoE framework, a gating network receives input data and generates a set of weights, typically in the form of a probability distribution, representing the selection probability for each expert model. This gating network dynamically selects the most appropriate expert model based on the current input data characteristics. Once the gating network has selected the relevant expert models, their outputs are combined through a fusion mechanism (such as weighted average or weighted sum) to produce the final model prediction.\\

For the FMCB detection task, we have conducted a series of different tests utilizing observational data acquired from various telescopes operating under different observational modes. Following extensive evaluation, we have identified two neural networks that exhibited superior performance in celestial object detection, each demonstrating distinct strengths across different detection scenarios. The first expert neural network (E1) is derived from the YOLO (You Only Look Once) architecture, while the second expert neural network (E2) is adapted from the CenterNet framework. Both of these neural networks have undergone modifications to incorporate the physical-inspired approach discussed in Section~\ref{sec2.1}, enhancing their ability to leverage domain-specific knowledge in the detection process. This adaptation aims to improve the networks' capacity to handle the unique challenges posed by FMCB. A detailed exposition of these neural network architectures, including their specific modifications and implementations, is presented in Section~\ref{sec2.3}. \\
    
\subsection{Detail Design of the Fast Moving Celestial Object Detection Neural Network} \label{sec2.3}

\begin{figure*}
  \centering
  \begin{minipage}{0.3\linewidth}
    \includegraphics[width=\linewidth]{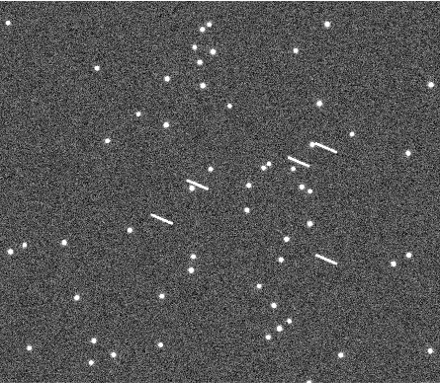} 
    \centering\footnotesize\textbf{(a)} 
  \end{minipage}%
  \hspace{0.05\linewidth}%
  \begin{minipage}{0.3\linewidth}
    \includegraphics[width=\linewidth]{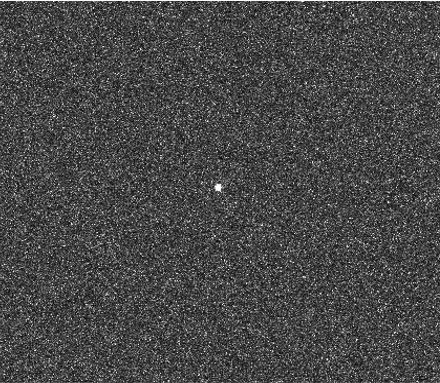}
    \centering\footnotesize\textbf{(b)} 
  \end{minipage}%
  \caption{Figure (a) shows the data in sidereal tracking mode, while Figure (b) shows the PSF convolved with motion vectors under the same sidereal tracking mode.}
  \label{figure_sample}
\end{figure*}

\begin{figure}
    \centering
    \includegraphics[scale=0.5]{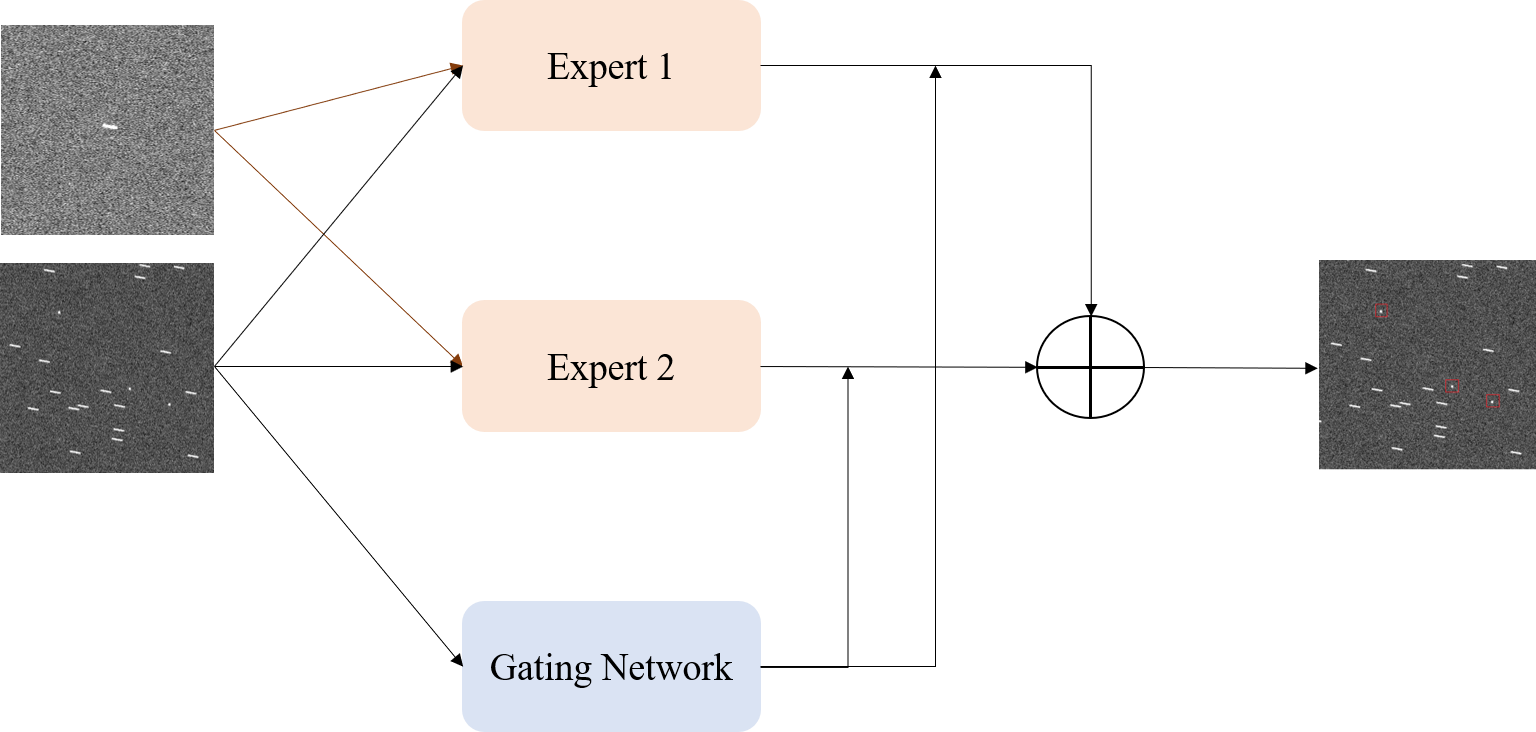}
    \caption{The structure of the MoE discussed in this paper. Two expert neural networks are used for FMCB detection, while one gating network is used to control these neural networks to output final detection results. Both of these two experts are PiNNs.The outputs of E1 and E2 are combined using the weights derived from the gating network to obtain our model.}
    \label{Figure2.3}
    \end{figure}

We propose utilizing contrastive learning to integrate detection results with images that simultaneously encode both the PSF and the observation mode. As described by \citet{chen2021semi}, contrastive learning is a machine learning approach aimed at extracting meaningful representations from data by emphasizing the similarities and differences between instances. However, traditionally contrastive learning often involves presenting positive and negative samples alongside the neural network during the training phase. In our approach, we present the negative samples (the PSF convolved with motion vectors) during both the training and deployment phases. During the training phase, the neural network will learn to manage detection results according to negative samples, while in the deployment phase, once the neural network is trained for processing real observational images, the negative samples will serve as references to filter out stars.
For example, the Figure \ref{figure_sample} shows an image  in sidereal tracking mode, where the targets (positive samples) appear as streaks, while the background stars appear as point-like shapes. Meanwhile, the PSF convolved with motion vectors appears as a point-like shape and is treated as a negative sample. This innovative application of contrastive learning may significantly improve the neural network's ability to identify and characterize FMCB by consistently leveraging the interaction between physical models and observational data. Based on our preceding discussion, we have designed a neural network specifically tailored for the detection of FMCB. The architecture of this neural network is illustrated in Figure \ref{Figure2.3}. As depicted, the network comprises three distinct components: Expert 1 (E1), Expert 2 (E2), and the gating neural network. It is worth noting that, in accordance with the MoE concept, our framework allows for the integration of additional detection neural networks, provided they are incorporated into the overall structure following the MoE principles. This modular design offers flexibility for future enhancements and adaptations to various detection scenarios. In the following sections, we will delve into the specifics of each component.\\


\subsubsection{Design of the Expert 1 Neural Network} \label{sec2.3.1}
\begin{figure}
    \centering
    \includegraphics[scale=0.5]{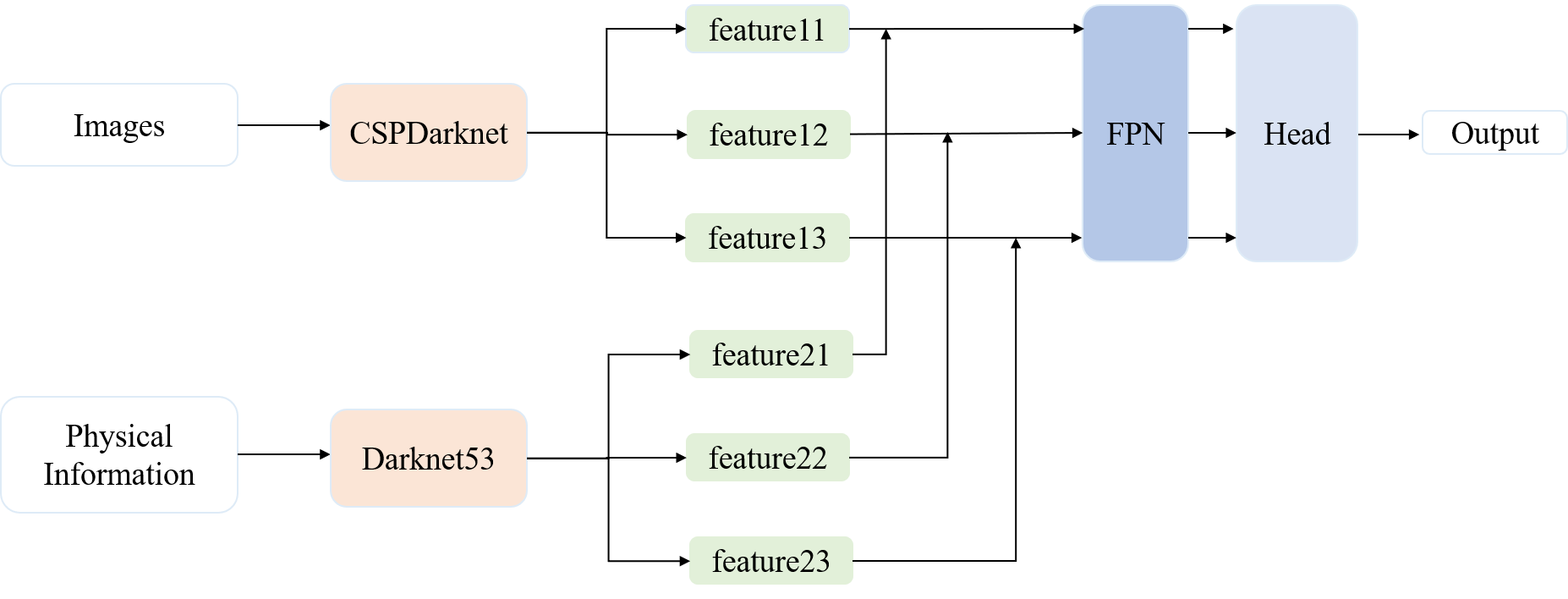}
    \caption{The Structure of the Neural Network Used in Expert 1.}
    \label{Figure expert1}
    \end{figure}

The structure of the E1 is shown in Figure~\ref{Figure expert1}. This network accepts two inputs: the observation image and the encoded image (which represents the convolution results of the PSF and the motion vector). E1's primary detection strategy is based on YOLOX \citep{ge2021yolox}, a single-step detection framework that provides three feature layers at different scales for FMCB detection. The observation image is processed by the detection component, while the encoded image is fed into a feature extraction neural network. In this study, we employ Darknet53 \citep{redmon2018yolov3} as the feature extraction neural network, which forms the core of YOLOv3 and has been widely adopted for object detection and classification tasks. As indicated in academic literature related to YOLOv3, DarkNet53 is favored for its straightforward design and ease of training. In practical applications, we have evaluated several different backbone neural networks, including ResNet50~\citep{he2016deep}, VGG16~\citep{simonyan2014very}, DarkNet19~\citep{redmon2017yolo9000}, and results have showed that DarkNet53 outperformed the other options. Therefore, we have chosen DarkNet53 for this study.  Features extracted by YOLOX from the observation image and those extracted by Darknet53 from the encoded image are concatenated, resulting in three fused features at different spatial scales. These fused features are then input into the Feature Pyramid Network (FPN), which performs feature fusion across different shapes of feature layers, facilitating more effective feature extraction \citep{lin2017feature}. The features processed by the FPN are subsequently passed to the YOLO Head to generate detection results.\\

In addition to the original loss function defined by YOLOX, which includes the localization loss \( \text{Loss}_{\text{reg}} \), the objectness loss \( \text{Loss}_{\text{obj}} \), the classification loss \( \text{Loss}_{\text{cls}} \), and the IoU loss \( \text{Loss}_{\text{reg}} \), we introduce a new loss function. This additional loss function is the cosine similarity between images of the detected object and the encoded image. In machine learning, cosine similarity is primarily used to measure the similarity between two sets of vectors in a two-dimensional or multidimensional coordinate system, checking whether they extend in the same direction within the same quadrant. The goal of this loss function is to maximize the cosine similarity of detected FMCB and the convolution result of the PSF and the moving vector (stars in star fields), thereby improving the model's ability to detect FMCB. Therefore, we incorporate cosine similarity loss as the contrastive loss and include it as part of the loss function in our FMCB detection network. All these loss functions are defined in Equation~\ref{formula1} ,where  \( N \) is the number of anchor points classified as positive samples.

\begin{equation}
\begin{split}
    \text{Loss} &= \frac{5 \cdot \text{Loss}_{\text{reg}} + \text{Loss}_{\text{cls}} + \text{Loss}_{\text{obj}} + \text{Loss}_{\text{cs}}}{N} \\
\end{split}
\label{formula1}
\end{equation}
\subsubsection{Design of the Expert 2 Neural Network} \label{sec2.3.2}
For our second expert network (E2), we choose CenterNet-based model due to its different detection strategy compared to the YOLOX-based model. The YOLOX-based model is primarily designed for FMCB detection in natural images, demonstrating proficiency in learning complex features to identify objects with diverse properties, while the CentetNet-based model is specifically developed for small FMCB detection. By incorporating these two contrasting approaches, we aim to capitalize on their respective strengths in detecting celestial objects across various astronomical scenarios. Unlike conventional object detection algorithms that rely on predefined anchor boxes, where numerous anchor boxes are distributed across the image and subsequently adjusted by the network to predict bounding boxes, the CenterNet-based model reframes the object detection problem as a center point prediction task. This innovative approach represents objects by their center points and utilizes predicted center point offsets along with width and height measurements to determine bounding boxes. The CenterNet-based methodology is notably streamlined, computationally efficient, and particularly well-suited for detecting small objects, leading to its widespread adoption across various domains \citep{zhou2019objects, ahmed2021edge, guo2021centernet++}. Our previous research has demonstrated the efficacy of an adapted CenterNet-based in detecting celestial objects for wide-field, small-aperture telescopes \citep{sun2023pnet}, further supporting its inclusion in our current framework. By integrating these complementary detection strategies within our MoE model, we anticipate enhanced performance across a broader range of astronomical detection scenarios, from densely populated star fields to sparse regions containing FMCB. \\

\begin{figure}
    \centering
    \includegraphics[scale=0.4]{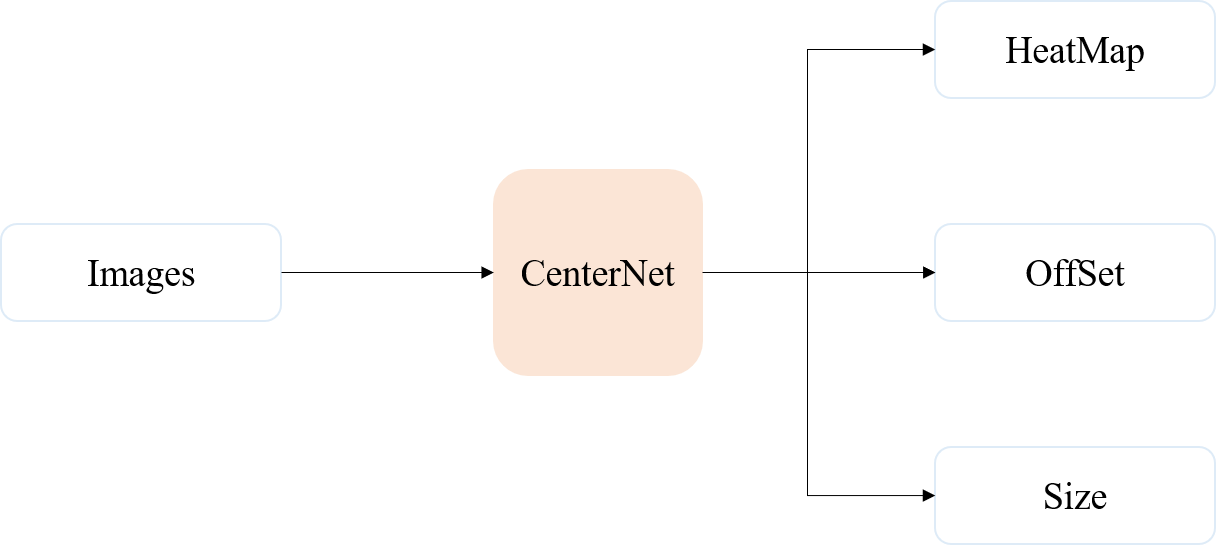}
    \caption{The Structure of the Neural Network Used in Expert 2.}
    \label{Figure expert2}
    \end{figure}

The architecture of E2 presented in this paper is illustrated in Figure~\ref{Figure expert2}. While various options exist for the CenterNet backbone neural network, including Hourglass \citep{newell2016stackedhourglassnetworkshuman}, DLA-Net \citep{su2021dlanetlearningduallocal}, and ResNet \citep{he2016deep}, we have chosen the classic ResNet50 as our primary feature extraction network. This selection is driven by ResNet50's 50-layer deep architecture, which allows for the learning of rich and complex feature representations. Moreover, its residual connections effectively alleviate the vanishing gradient problem, promoting faster convergence of the network. In our implementation, the input observation image undergoes processing through the ResNet50 to achieve comprehensive feature extraction.\\

Our approach to integrating detection results with physical-inspired information employs a unique strategy. The CenterNet model generates three primary outputs: a heatmap, center point offsets, and width/height predictions. As a result, our loss function consists of three essential components: heatmap loss, width/height loss, and center point regression loss. The heatmap loss, represented as $\text{Loss}_{\text{c}}$, measures the accuracy of predicted object center points. For each object category, the model produces a heatmap that illustrates the probability distribution of the object's center point across all image pixels. The width/height loss, $\text{Loss}_{\text{wh}}$, and center point regression loss, $\text{Loss}_{\text{off}}$, are calculated using L1 loss, which ensures both accurate predictions of the object's bounding box dimensions and precise localization of the object center point. To improve the model's capability to differentiate celestial objects based on their physical characteristics, we introduce an additional component: the cosine similarity loss, $\text{Loss}_{\text{cs}}$. This term evaluates the similarity between the detected object's image and the PSF, thereby incorporating physical based knowledge into the detection process. The total loss function is defined as Equation \ref{formula2},
\begin{align}
    \text{Loss} &= \text{Loss}_{\text{c}} + 0.1 \cdot \text{Loss}_{\text{wh}} + \text{Loss}_{\text{off}} + \text{Loss}_{\text{cs}}.
    \label{formula2}
\end{align}
The architecture of E2, as described above, enables us to effectively detect celestial objects in observation images while simultaneously distinguishing potential FMCB of interest from stationary stars.\\

\subsubsection{Design of the Gating Network in the MoE}
\label{sec2.3.3}
\begin{figure}
    \centering
    \includegraphics[scale=0.4]{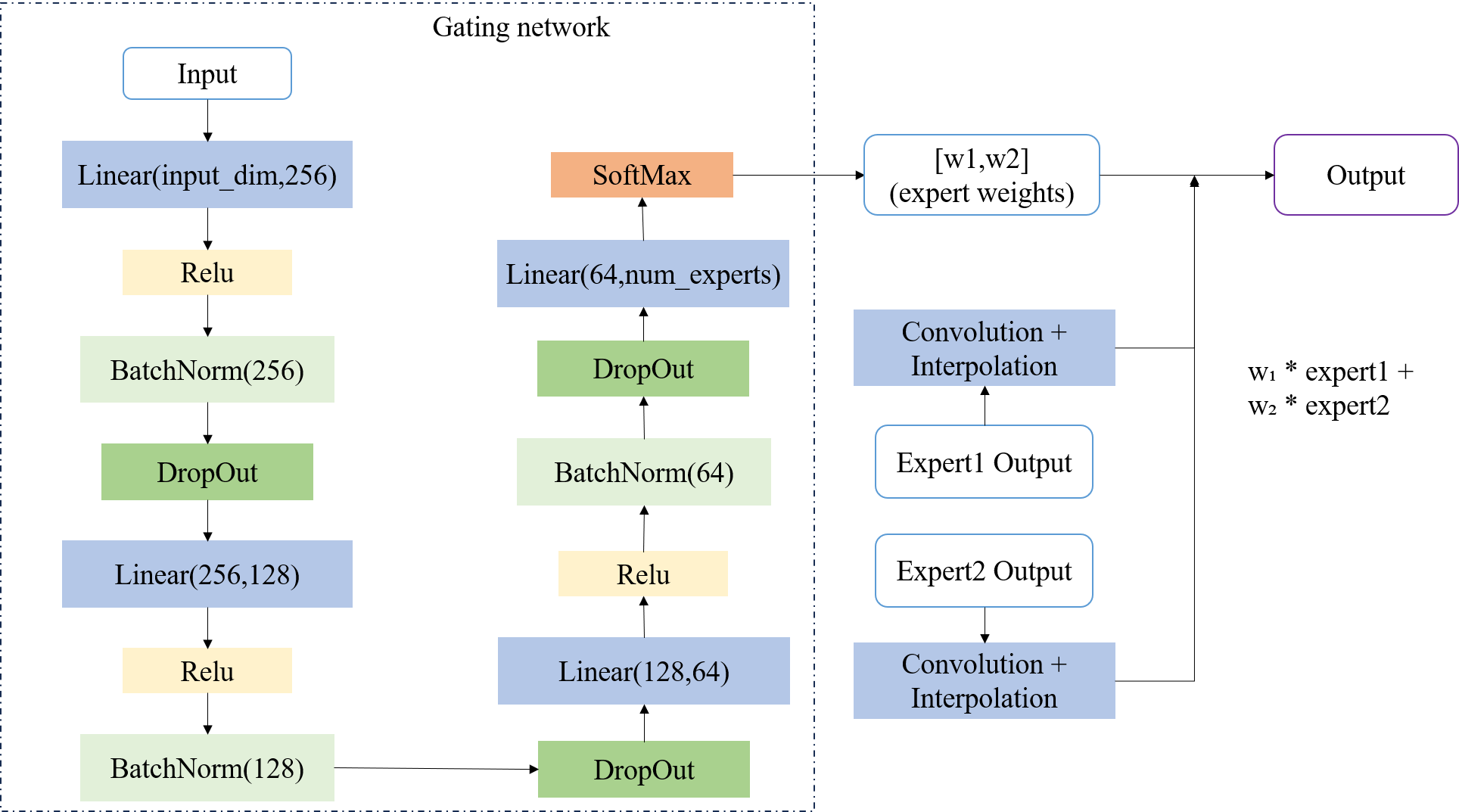}
    \caption{The structure of the gating network.}
    \label{Figuregate}
    \end{figure}

The MoE is employed to synthesize results from multiple expert networks, ultimately producing a final output. While various methods exist for constructing an MoE system, we have adopted the widely-used gating network architecture for our implementation \citep{eigen2013learning}. In our design, the gating network accepts the observation image as input, which is initially flattened into a one-dimensional vector to conform to the input requirements of the fully connected layers within the network. Figure~\ref{Figuregate} illustrates the structure of our gating network. It comprises a series of processing stages, including multiple linear layers, activation functions, batch normalization, and dropout layers. This sequence of operations culminates in the application of a softmax function, which transforms the output of the final linear layer into a probability distribution. This ensures that the network's output represents a valid set of weights for combining expert predictions. The gating network's role is crucial in determining the contribution of each expert to the final result. It dynamically assigns weights to the outputs of individual expert networks based on the input image characteristics. The final detection results are then obtained by multiplying each expert's output with its corresponding weight derived from the gating network. This adaptive weighting mechanism allows our system to leverage the strengths of different experts across varying observational scenarios, potentially enhancing overall detection performance and robustness.\\

The gating network, as described in Figure~\ref{Figuregate}, serves as the cornerstone for integrating results from our expert networks within the MoE framework. The gating network takes a flattened representation of the input, specifically, a three-channel image with dimensions of $512 \times 512$, which is converted into a one-dimensional vector of length $3 \times 512 \times 512$ . This vector is subsequently processed through a series of fully connected layers, activation functions, batch normalization, and dropout layers to yield a two-dimensional output. The output from the gating network indicates a set of weights that are assigned to each expert neural network, culminating in a final layer that utilizes a softmax function to generate a probability distribution over all experts. At the same time, each expert network processes the input independently, producing unique detection results. However, these individual outputs vary in spatial dimensions and channel sizes. To ensure compatibility among these outputs, a series of refinement processes are employed, including convolutional operations to adjust channel dimensions and interpolation methods to align spatial dimensions. These operations enable the feature maps generated by the experts to be resized consistently, thereby ensuring they can be effectively combined during the weighted aggregation step. This dual-weight system allows for dynamic balancing of contributions from both experts based on input characteristics. Concurrently, each expert network generates its own set of detection results. The final step in our MoE pipeline involves the weighted combination of these processed expert outputs, utilizing the weights derived from the gating network. \\

\section{Training and Validation of the Algorithm} \label{3}
To effectively illustrate the capabilities of our methodology, we will generate a specific collection of simulation images to serve as training data, encompassing three distinct observation modes, although not all conceivable observation scenarios. Subsequent to training the neural network, we will produce new sets of observation images that also incorporate these three observation modes but utilize completely different parameter sets. All tests will be conducted on a computer featuring an NVIDIA GeForce RTX 3090 GPU, an Intel(R) Xeon(R) Gold 5218 CPU, and 512GB of memory. We will begin with a thorough discussion of the training and validation processes associated with our algorithms. In addition, we will design two observation scenarios to demonstrate the performance of our framework in processing data that varies with different sky backgrounds and observation modes. At last, we show the performance of our algorithm with real observation images obtained by two different observation modes.\\

\subsection{Data Generation and Scenario Definition}\label{3.1}
Since the observation of FMCB involves three distinct observation modes, each encompassing numerous scenarios, acquiring real observational data that could represent these situations becomes quite challenging. This complexity underlies our proposal for a framework to detect FMCB. To more effectively demonstrate the capabilities of our neural network, we utilize simulated images to train and evaluate our algorithm, as simulation techniques can generate images tailored to our specific requirements. We employ the Skymaker for the generation of various simulated images. There are three distinct observation modes defined in this paper. Observation with Tracking Mode: the telescope follows FMCB, resulting in streaked images of stars and point-like images of the FMCB. Observation with Sky Survey Mode: the telescope tracks stars, leading to point-like images of the stars, while FMCB appear as streaked images. Observation with Intersection Mode: the telescope moves in arbitrary directions, causing both stars and FMCB to be captured as streaked images, varying in length and orientation.\\

\begin{figure}
\centering
\begin{minipage}[b]{0.3\textwidth}
    \centering
    \includegraphics[width=90pt,height=90pt]{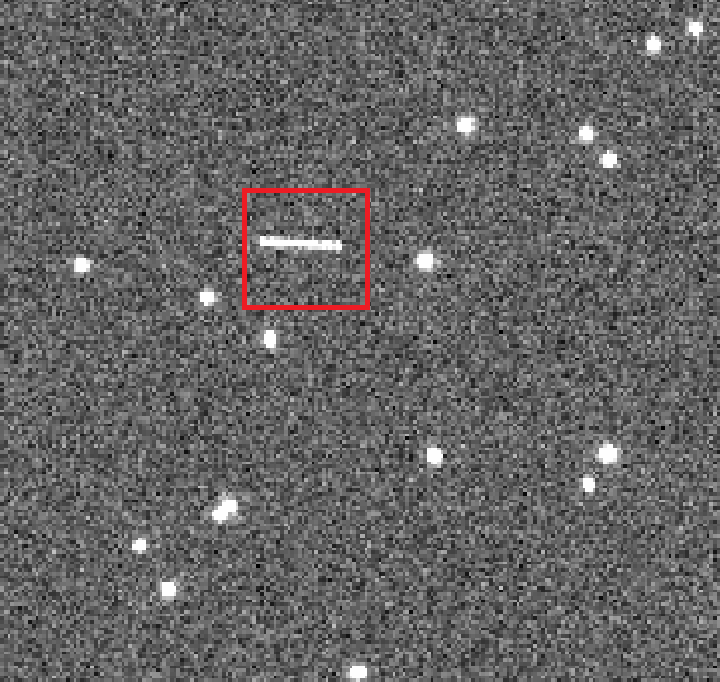}
    \label{fig:tracking}
\end{minipage}
\quad
\begin{minipage}[b]{0.3\textwidth}
    \centering
    \includegraphics[width=90pt,height=90pt]{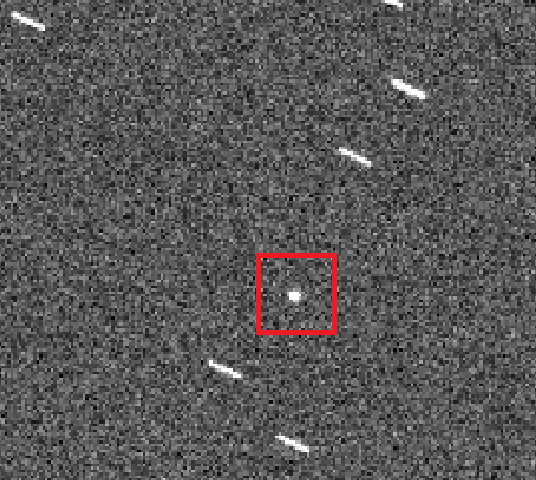}
    \label{fig:following} 
\end{minipage}
\quad
\begin{minipage}[b]{0.3\textwidth}
    \centering
    \includegraphics[width=90pt,height=90pt]{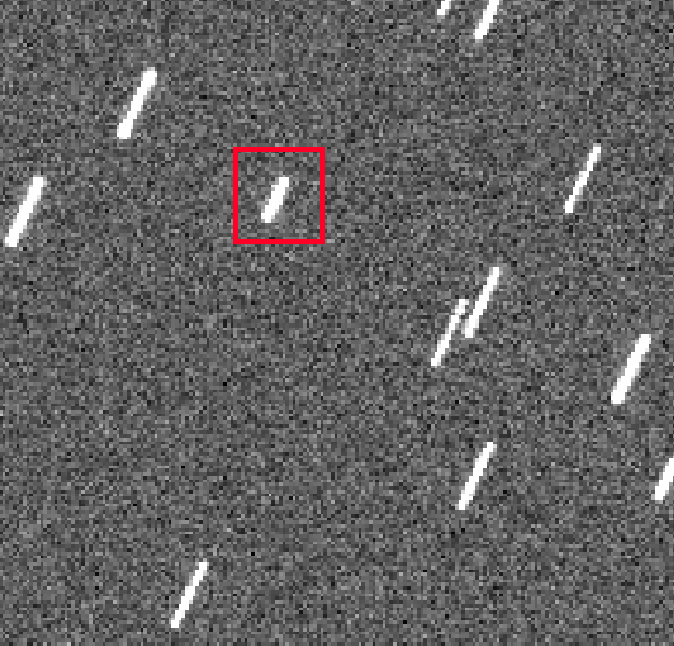}
    \label{fig:intersection}
\end{minipage}
\caption{This figure presents images corresponding to three distinct observation modes. From left to right, the images represent observations made in sidereal mode, tracking mode, and intersection mode, respectively. All images are captured under optimal observation conditions, characterized by  a sky background of 16.9. For enhanced clarity and ease of identification, the FMCB are highlighted with red bounding boxes in each image.}
\label{fig:all_modes}
\end{figure}

In these simulations, we use the following steps to create images of background stars and FMCB. Firstly, we generate pre-images, which are images virtually observed by the telescope with very short exposure time (1/100 second to 1/10 second) and no read-out noise, ensuring that the background stars and FMCB remain stationary relative to the telescope's pointing direction. Positions and magnitudes of stars and moving celestial objects in each sub-images are calculated according to the observation mode, pixel scale of the telescope, and exposure duration. Then all these sub-images will be stacked together and we add read-out noise into stacked images. If the exposure time of one observation image is divided into M intervals, we generate M frames of pre-images using the Skymaker. Since the FMCB detection algorithm needs position of celestial objects as label, we will set the geometric center of all FMCB as position of these FMCB. In Figure~\ref{fig:all_modes}, we present three frames observational images obtained by different observation modes. In this paper, we consider variations of observation conditions, including the sky background. We define the sky background level as a random variable between 14.0 and 19.0. In addition, we consider the direction of movement of a fast moving object relative to the telescope as a random variable from $0 \deg$ to $180 \deg$, with a relative velocity from and 0 to 3.38 pixels per frame. All of these simulation images are generated with a pixel scale of 0.900 arcseconds, and the PSF in the observation images is modeled as a Gaussian function with a full width at half maximum (FWHM) ranging from 0.9 to 3 arcseconds. These parameters are established with consideration for the misalignment and defocus inherent in wide field telescopes, which are generally used for space-based observations. \\

\subsection{Performance Evaluation Criteria}\label{3.2}
To evaluate the model's performance, this paper uses the mean Average Precision (mAP) at an Intersection over Union (IoU) threshold of 0.5 as the primary metric, along with a comprehensive analysis of precision and recall. IoU is defined as the ratio of the intersection area to the union area of the predicted and ground truth bounding boxes, as shown in Equation~\ref{formula3}:
\begin{equation}
    \text{IoU} = \frac{|A \cap B|}{|A \cup B|},
    \label{formula3}
\end{equation}
where \( |A \cap B| \) is the area of intersection between sets \( A \) and \( B \), and \( |A \cup B| \) is the area of their union. Precision measures the proportion of correctly predicted positive samples out of all samples predicted as positive, which is defined in Equation~\ref{formula4}:
\begin{equation}
    \text{Precision} = \frac{TP}{TP + FP},
    \label{formula4}
\end{equation}
where \( TP \) is the number of true positives and \( FP \) is the number of false positives in all detection results. Recall measures the proportion of correctly predicted positive samples out of all actual positive samples, which is defined in Equation~\ref{formula5}:
\begin{equation}
    \text{Recall} = \frac{TP}{TP + FN},
    \label{formula5}
\end{equation}
where \( TP \) is the number of true positives and \( FN \) is the number of false negatives. For object detection tasks, Precision and Recall can be calculated for each class. Each class will have its own Precision-Recall (P-R) curve, and the area under this curve is the Average Precision (AP) for that class. The mean Average Precision (mAP) is the average of the AP values across all classes, which is defined in Equation~\ref{formula6}:
\begin{equation}
    \text{mAP} = \frac{1}{n} \sum_{i=1}^{n} \text{AP}_i,
    \label{formula6}
\end{equation}
where \( n \) is the number of classes, and \( \text{AP}_i \) is the average precision for class \( c_i \). The mAP metric offers an easy way to evaluate the performance of object detection algorithms. It not only strikes a crucial balance between precision and recall rates but also provides an efficient and robust evaluation criterion that circumvents the need for complex threshold adjustments. This comprehensive approach allows for a more nuanced assessment of the performance of the detection algorithm across varying confidence levels, effectively capturing the algorithm's ability to identify objects accurately while minimizing false positives. Therefore, we use the mAP to evaluate the performance of our algorithm.\\

\subsection{Training of the Neural Network}\label{3.3}
Our neural network is designed to generalize well across various scenarios without requiring exhaustive training data. To evaluate and train our neural network, we only generate a dataset comprising 3000 simulated images, evenly distributed across three observational modes (1000 images per mode). Each image is accompanied by its corresponding PSF, motion vector, and the precise position of the FMCB, serving as prior information and ground truth labels. We first divide the dataset into training and validation sets in a 7:3 ratio. To expedite model convergence, we leverage pre-trained weights provided by PyTorch for initializing the CSPDarknet and ResNet50 networks, which are responsible for feature extraction. Subsequently,to enhance the model's generalization capabilities and mitigate overfitting, we implement a suite of data augmentation techniques. These include random rotations, horizontal flips, random cropping, and mosaic data augmentation, collectively expanding the diversity of our training samples. The data pipeline, including loading and batching operations, is managed using PyTorch's Dataset and DataLoader classes, ensuring efficient and consistent data handling throughout the training process.\\

Given the complex architecture of our MoE model, comprising several distinct neural networks, we adopt a multi-stage training strategy. Initially, we load pre-trained weights and train each constituent neural network independently. Once individual networks achieved convergence, we proceed to fine-tune the entire MoE system. It is important to highlight the distinctions between the frozen and unfrozen training phases, as they serve different purposes in our model optimization strategy. During the frozen training phase, we freeze weights of the model's backbone, preventing updates to the feature extraction network. This approach offers two key advantages: it significantly reduces GPU memory consumption and allows for fast moving celestial objected fine-tuning of specific network components. Conversely, in the unfrozen training phase, we release all constraints on the model's backbone, enabling comprehensive updates to the feature extraction network. While this comprehensive training approach necessitates increased GPU memory usage due to the adjustment of all network parameters, it allows for more thorough optimization.  This two-stage training protocol thus balances the need for rapid initial convergence with the requirement for comprehensive model refinement, ultimately contributing to a more robust and efficient training process for our complex neural network architecture. Outliers in the training dataset can cause fluctuations in the loss curve across epochs, complicating the assessment of performance variations. To address this issue, we will employ the Savitzky-Golay filter~\citep{5888646} to smooth the loss curve. This filtering technique effectively diminishes high-frequency noise and irregularities, leading to more stable and uniform curves. Consequently, it enhances the clarity with which the model's learning progress can be observed.\\

\begin{figure}
    \centering
    \begin{minipage}{0.3\textwidth}
        \centering
        \includegraphics[width=\linewidth]{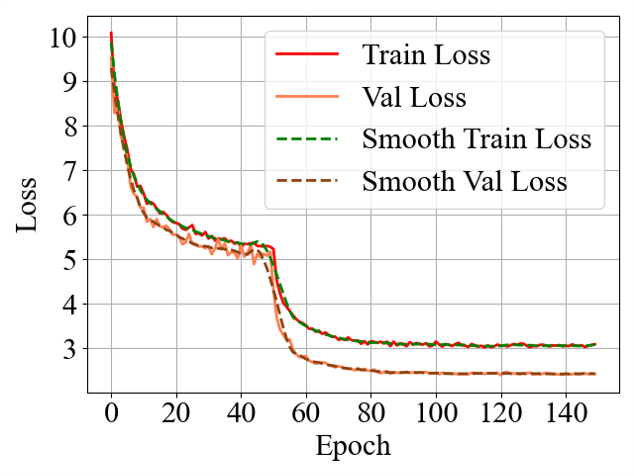}
        \centering\footnotesize\textbf{(a)}
    \end{minipage}%
    \hspace{0.03\textwidth}%
    \begin{minipage}{0.3\textwidth}
        \centering
        \includegraphics[width=\linewidth]{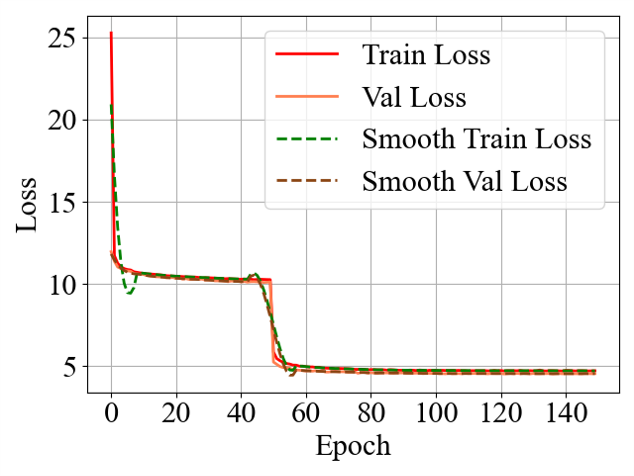}
        \centering\footnotesize\textbf{(b)} 
    \end{minipage}%
    \hspace{0.03\textwidth}%
    \begin{minipage}{0.3\textwidth}
        \centering
        \includegraphics[width=\linewidth]{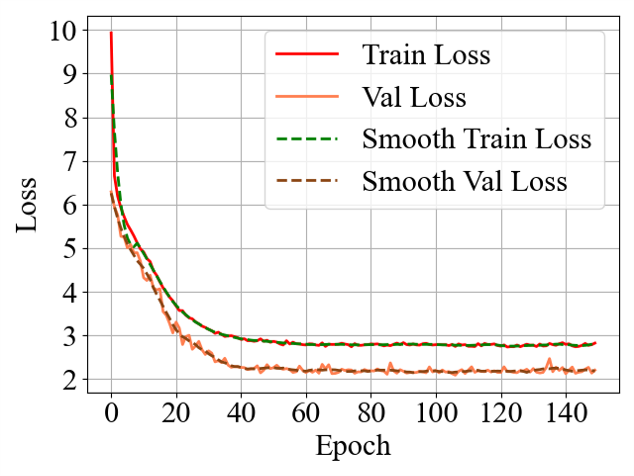}
        \centering\footnotesize\textbf{(c)} 
    \end{minipage}
    \caption{Evolution of loss values during the training process. Figure (a) illustrates the loss values for E1, Figure (b) illustrates the loss values for E2, and Figure (c) illustrates the loss values for the MoE. These figures demonstrate that both the individual Expert neural networks and the integrated MoE system achieve convergence within a span of tens to hundreds of epochs.}
    \label{Figure5}
\end{figure}

For the E1 neural network, the loss curve is shown in the left panel of Figure~\ref{Figure5}. It costs 150 epochs to train the neural network, with the first 50 epochs dedicated to the frozen phase. During this initial stage, we employ a learning rate of 0.001 and a batch size of 16. Transitioning to the unfrozen phase, we reduce the learning rate to 0.00001 and adjust the batch size to 8 to accommodate the increased computational demands of full model training. Throughout the process, we utilize 4 worker threads for data loading and employ the Adam optimizer with a weight decay of 0.0005. Learning rate adjustments are managed using the StepLR scheduler. As training progressed, we observe a consistent decrease in the loss function. By the 100th epoch, the loss stabilize, indicating model convergence. The entire training process for E1 require approximately 11 hours on our computer.\\

The training protocol for E2 mirrors that of E1, employing a two-phase approach: frozen and unfrozen training. We allocate a total of 150 epochs for the entire process, with the initial 50 epochs dedicated to the frozen phase, as shown in the middle panel of Figure~\ref{Figure5}. During this stage, we utilize an initial learning rate of 0.001 and a batch size of 16. Transitioning to the unfrozen phase, we adjust the learning rate to 0.00001 and reduce the batch size to 8 to accommodate the increased computational demands of full model training. Throughout the training process, we maintain consistent parameters, including 4 worker threads for data loading and the Adam optimizer with a weight decay of 0.0005. Learning rate adjustments were managed dynamically using the StepLR scheduler. As the training progressed, we observe a steady decline in the loss function. By the 100th epoch, the loss stabilized, signaling model convergence. The complete training cycle for E2 required approximately 9 hours on our computational infrastructure.\\

The training procedure for the MoE encompasses a comprehensive approach to optimization. As shown in the right panel of Figure~\ref{Figure5}, we employ a training regimen consisting of 150 epochs, initializing with a learning rate of 0.00001 and utilizing a batch size of 4. To facilitate efficient data processing, we allocate 4 worker threads. The Adam optimizer is selected for its adaptive learning rate capabilities, while the StepLR mechanism is implemented to manage learning rate adjustments dynamically throughout the training process. As the training progressed, we observe a plateauing of the loss function around the 100th epoch, indicating model convergence. The entire training cycle for the MoE require approximately 19 hours on our computational infrastructure. To mitigate the risk of overfitting, we select the model state at the 100th epoch as our final model, striking a balance between performance and generalization.\\

\subsection{Performance Evaluation with the Validation Set}\label{3.4}
\subsubsection{Overall Performance Evaluation with the Validation Set} \label{3.4.1}
\begin{figure}[ht]
    \centering
    \includegraphics[scale=0.5]{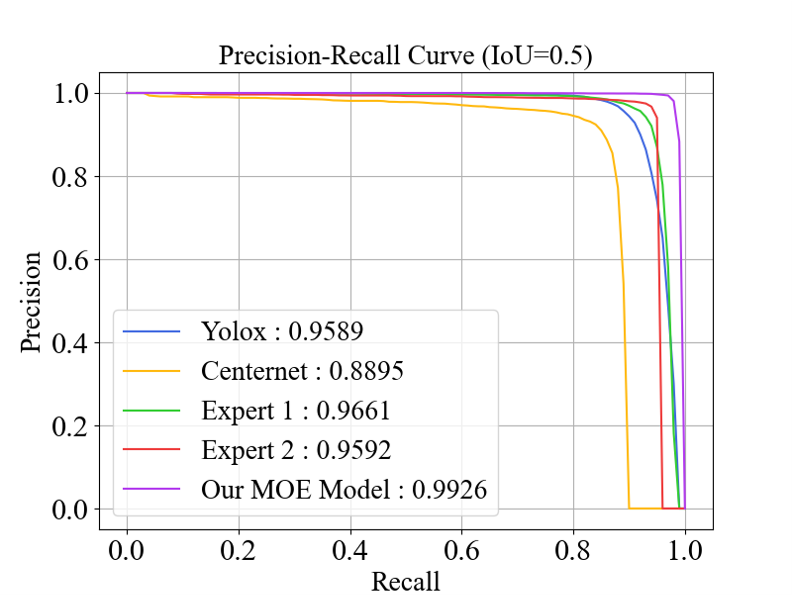}
    \caption{Precision-Recall (PR) Curve for different detection algorithms in the validation set with IoU of 0.5.}
    \label{fig:pr_curve}
\end{figure}

Following the training phase, we have initially employed our validation set to assess the performance of our MoE model. For comparative analysis, we have also adapted vanilla YOLOX and CenterNet for the detection of FMCB, where "vanilla YOLOX" and "vanilla CenterNet" refer to the unmodified, original versions of the YOLOX and CenterNet neural networks, respectively, as these frameworks represent state-of-the-art performance according to previous literature. Our MoE model has demonstrated exceptional performance, achieving a mean Average Precision (mAP) of $99.26\%$, a precision rate of $98.50\%$, and a recall rate of $98.57\%$. In contrast, other detection methods exhibited comparatively lower mAP scores, as illustrated in Table~\ref{table1} and Figure~\ref{fig:pr_curve}. The precision-recall curve in Figure~\ref{fig:pr_curve} clearly demonstrates that the MoE model outperforms all other tested approaches. Notably, both E1 and E2 components of our MoE system show superior performance compared to their vanilla neural network architectures. This improvement suggests that the incorporation of physical-inspired information enhances the detection capabilities for FMCB, underscoring the efficacy of our approach in this specialized astronomical application.\\

\begin{table}[ht]
    \centering
    \caption{Comparative analysis of the performance of different detection algorithms in processing images in the validation set.}
    \label{table1}
    \begin{tabular}{|l|c|c|c|}
        \hline
        Method & mAP & Precision & Recall \\
        \hline
        YOLOX & 95.89\% & 75.61\% & 94.80\% \\
        CenterNet & 88.95\% & 98.06\% & 62.94\% \\
        Expert1 & 96.61\% & 91.39\% & 94.64\% \\
        Expert2 & 95.92\% & 99.07\% & 82.40\% \\
        Our MoE Model & 99.26\% & 98.50\% & 98.57\% \\
        \hline
    \end{tabular}
\end{table}

\begin{figure}
    \centering
    \includegraphics[scale=0.5]{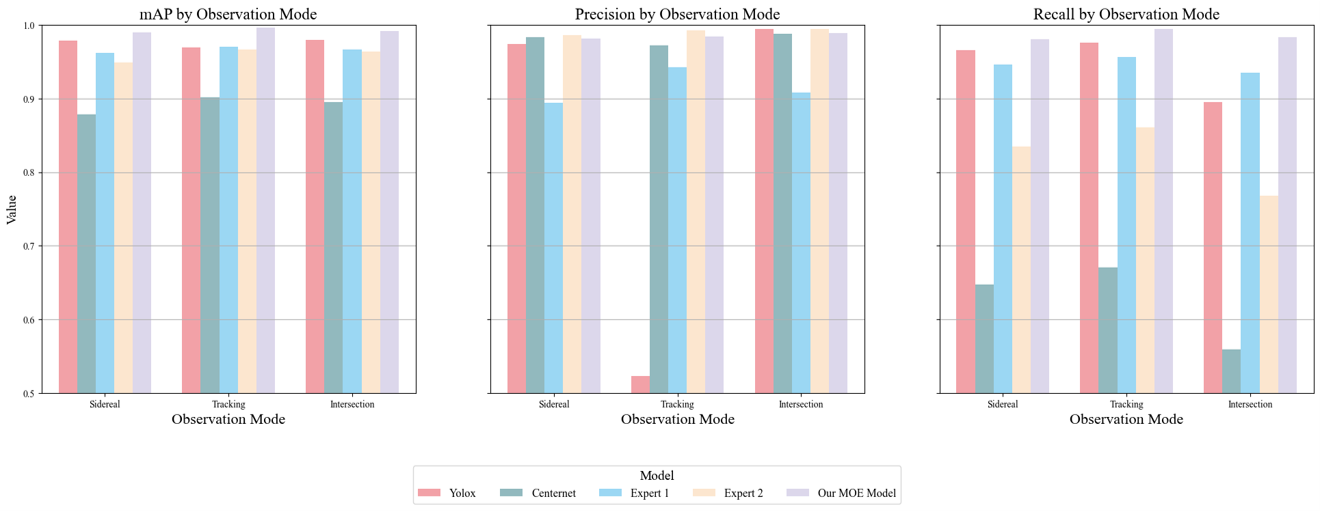}
    \caption{mAP, Precision, and Recall for each model under different observation modes}
    \label{Figure7}
\end{figure}

\begin{figure}
\centering
\begin{minipage}[b]{0.3\textwidth}
    \centering
    \includegraphics[width=\textwidth]{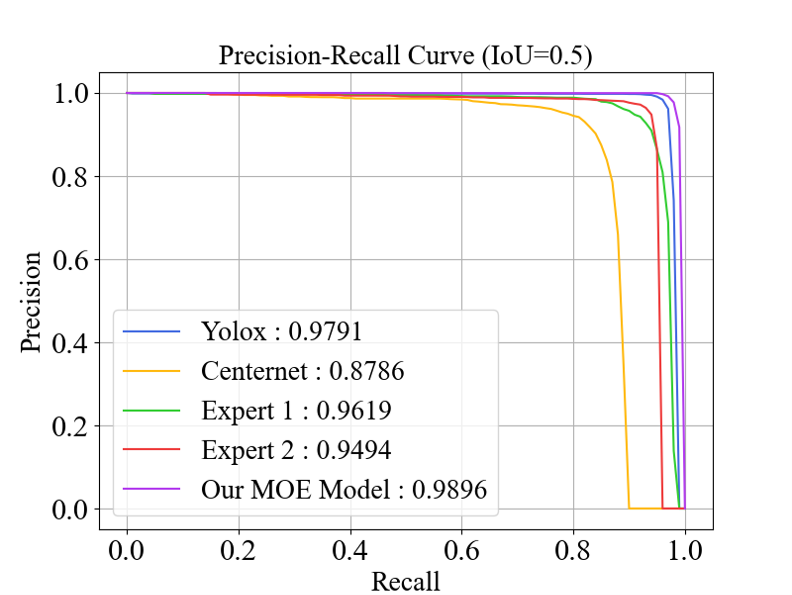}
\end{minipage}
\vspace{0.5cm} 
\begin{minipage}[b]{0.3\textwidth}
    \centering
    \includegraphics[width=\textwidth]{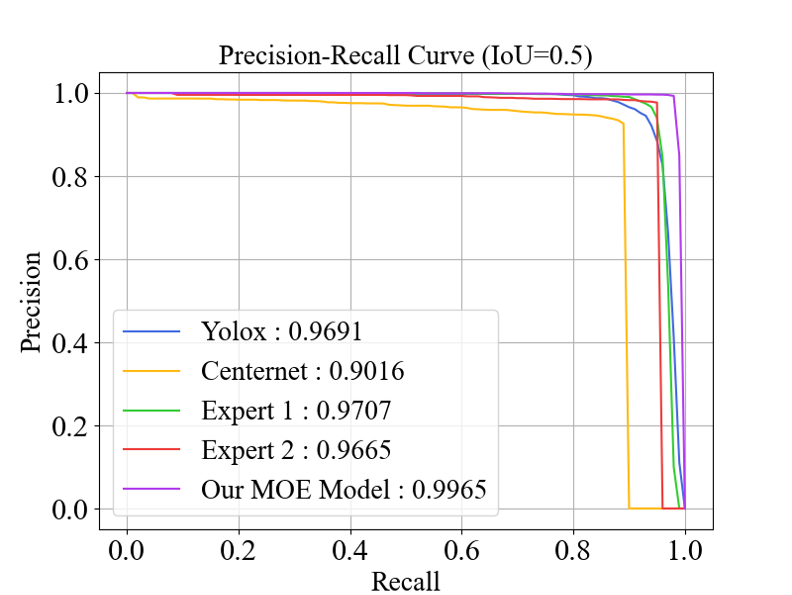}
\end{minipage}
\vspace{0.5cm} 
\begin{minipage}[b]{0.3\textwidth}
    \centering
    \includegraphics[width=\textwidth]{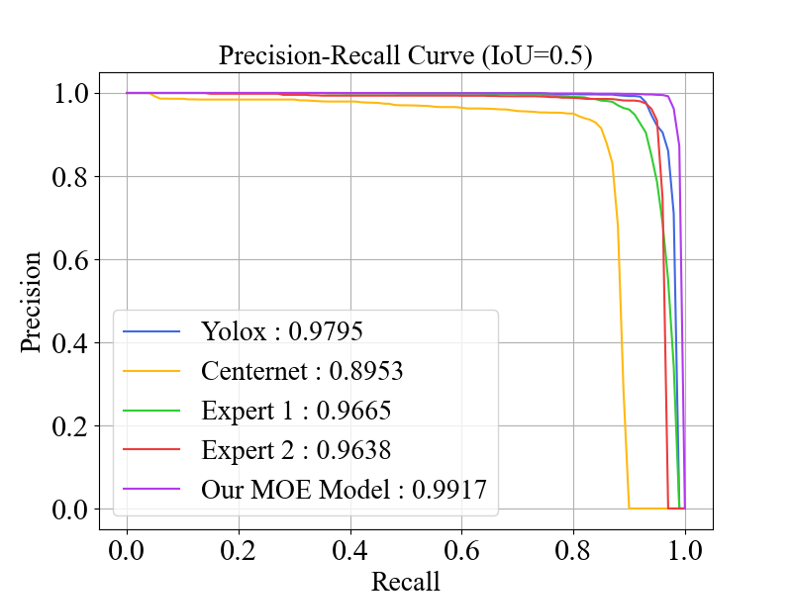}
\end{minipage}
\caption{The Precision-Recall (PR) curves for each model under different observation modes}
\label{Figure8}
\end{figure}

Among the three observational modes, we anticipate that the intersection mode would present the greatest challenge due to its encompassing of diverse observational scenarios. In these scenarios, the morphology of stars and FMCB can vary significantly, posing substantial challenges for source detection algorithms. Consequently, it is imperative to evaluate our algorithm's performance separately with images obtained by three different observational modes, which could help us to better understand the performance variations of our methods in different applications. Table~\ref{table2} presents the detection results for each observational mode, corroborating our initial assumptions. Vanilla deep learning-based FMCB detection algorithms exhibit the poorest performance in the intersection mode. In contrast, E1 and E2 demonstrate superior performance compared to classical methods, benefiting from the integration of physical-inspired information. The MoE model, which combines E1 and E2, achieves the best performance in FMCB detection across all three observational modes. Figure~\ref{Figure7} and Figure~\ref{Figure8} provide a more intuitive illustration of these results. Analysis of the overall detection performance on the validation dataset reveals that our method attains the highest mAP value and demonstrates superior performance in both Precision and Recall metrics. Furthermore, our approach achieves the best mAP values across all observational modes, consistently exhibiting the highest precision and recall rates. In conclusion, our method demonstrates superior overall performance, effectively addressing the challenges posed by diverse observational scenarios and outperforming existing approaches in the detection of FMCB. To further validate the performance of our model, we test the model using 1500 simulated images with parameters that are unrelated to the training data, with each observation mode consisting of 500 images. The detection results are shown in the Table ~\ref{table4} and Table~\ref{table5}. The results are consistent with the results obtained by the validation set, and our method achieved the best performance.\\

\begin{table}[ht]
    \centering
    \caption{Comparative analysis of the performance of different detection algorithms in processing images obtained by three different observational modes.}
    \label{table2}
    \begin{tabular}{|l|c|c|c|c|c|c|c|c|c|}
        \hline
        \textbf{Method} & \multicolumn{3}{c|}{\textbf{Sidereal Mode}} & \multicolumn{3}{c|}{\textbf{Tracking Mode}} & \multicolumn{3}{c|}{\textbf{Intersection Mode}} \\
        \cline{2-10}
        & \textbf{mAP} & \textbf{Precision} & \textbf{Recall} & \textbf{mAP} & \textbf{Precision} & \textbf{Recall} & \textbf{mAP} & \textbf{Precision} & \textbf{Recall} \\
        \hline
        YOLOX & 97.91\% & 97.40\% & 96.54\% & 96.91\% & 52.34\% & 97.58\% & 97.95\% & 99.41\% & 89.49\% \\
        CenterNet & 87.86\% & 98.37\% & 64.78\% & 90.16\% & 97.18\% & 67.08\% & 89.53\% & 98.80\% & 55.96\% \\
        Expert 1 & 96.19\% & 89.40\% & 94.62\% & 97.07\% & 94.22\% & 95.68\% & 96.65\% & 90.82\% & 93.51\% \\
        Expert 2 & 94.94\% & 98.64\% & 83.49\% & 96.65\% & 99.25\% & 86.13\% & 96.38\% & 99.44\% & 76.84\% \\
        Our MoE Model & 98.96\% & 98.18\% & 98.01\% & 99.65\% & 98.40\% & 99.41\% & 99.17\% & 98.92\% & 98.32\% \\
        \hline
    \end{tabular}
\end{table}

\begin{table}
    \centering
    \caption{Comparative analysis of the performance of different detection algorithms in processing images in a new dataset. In the dataset, images are generated with new sets of parameters that are randomly generated.}
    \label{table4}
    \begin{tabular}{|l|c|c|c|}
        \hline
        \textbf{Method} & \textbf{mAP} & \textbf{Precision} & \textbf{Recall} \\
        \hline
        YOLOX & 97.01\% & 75.85\% & 96.13\% \\
        CenterNet & 84.73\% & 95.86\% & 63.81\% \\
        Expert1 & 97.42\% & 92.04\% & 95.89\% \\
        Expert2 & 96.63\% & 99.23\% & 84.61\% \\
        Our MoE Model & 99.48\% & 98.31\% & 98.84\% \\
        \hline
    \end{tabular}
\end{table}

\begin{table}
    \centering
    \caption{Comparative analysis of the performance of different detection algorithms in processing images obtained by three different observational modes. In the dataset, images are generated with new sets of parameters that are randomly generated.}
    \label{table5}
    \begin{tabular}{|l|c|c|c|c|c|c|c|c|c|}
        \hline
        \textbf{Method} & \multicolumn{3}{c|}{\textbf{Sidereal Mode}} & \multicolumn{3}{c|}{\textbf{Tracking Mode}} & \multicolumn{3}{c|}{\textbf{Intersection Mode}} \\
        \cline{2-10}
        & \textbf{mAP} & \textbf{Precision} & \textbf{Recall} & \textbf{mAP} & \textbf{Precision} & \textbf{Recall} & \textbf{mAP} & \textbf{Precision} & \textbf{Recall} \\
        \hline
        YOLOX & 99.57\% & 98.55\% & 98.42\% & 97.11\% & 52.10\% & 97.98\% & 97.91\% & 99.42\% & 91.73\% \\
        CenterNet & 79.94\% & 93.13\% & 65.01\% & 90.16\% & 97.40\% & 66.80\% & 85.28\% & 97.38\% & 59.51\% \\
        Expert 1 & 98.03\% & 90.54\% & 96.96\% & 97.68\% & 94.20\% & 96.83\% & 96.59\% & 91.47\% & 93.84\% \\
        Expert 2 & 95.98\% & 98.65\% & 85.86\% & 97.36\% & 99.40\% & 87.13\% & 96.69\% & 99.70\% & 80.73\% \\
        Our MoE Model & 99.70\% & 98.09\% & 99.13\% & 99.67\% & 98.72\% & 99.37\% & 99.05\% & 98.01\% & 98.03\% \\
        \hline
    \end{tabular}
\end{table}

\subsubsection{Performance Comparison between Different Experts} \label{3.4.2}
According to the results obtained from the validation set, we can find that E1 demonstrates slightly better performance due to the introduction of the physically inspired neural network. However, the performance of E2 surpasses that of CenterNet significantly, warranting further investigation into this phenomenon. In our experience, CenterNet exhibits high sensitivity to variations in sky background \citep{sun2023pnet}. Consequently, we have designed an additional test to evaluate the robustness that the physically inspired neural network provides for CenterNet. We have generated a dataset featuring sky backgrounds ranging from 14.0 to 19.9 across five different levels ,the values decreasing from level 1 to 5, and have subsequently analyzed the recall rates of the data under various background values, as illustrated in Table~\ref{table3.4.2}.\\.

\begin{table}[ht]
\centering
\caption{Recall rate of CenterNet and Expert2 under images with different background levels}
\label{table3.4.2}
\begin{tabular}{|c|c|c|c|c|c|}
\hline
\textbf{Sky Background (mag)} & 19.9 & 18.425 & 16.950 & 15.475 & 14 \\
\hline
\textbf{CenterNet recall} & 46.58\% & 27.16\% & 13.16\% & 1.17\% & 0\% \\

\textbf{Expert2 recall} & 70.44\% & 55.27\% & 41.95\% & 27.76\% & 14.01\% \\
\hline
\end{tabular}
\end{table}

From Table~\ref{table3.4.2}, it can be seen that both CenterNet and Expert2 perform poorly with bright backgrounds. This is because a bright background increases background noise, reducing the difference between the FMCB and the background, which affects the model's detection ability. Specifically, CenterNet's recall rate drops to 0\% under the brightest background, indicating that CenterNet is very sensitive to background noise and is unable to effectively recognize celestial objects. Although Expert2's recall rate also decreases as the background brightness increases, it still maintains a certain level of recall (14.01\%) under bright backgrounds. This suggests that Expert2 is more adaptable to brighter backgrounds and has slightly stronger capabilities in feature extraction and FMCB recognition. Besides, it should be mentioned that although CenterNet and Expert2 are inferior to Expert1 in terms of object detection capability, the goal of using two experts is to leverage the advantages of the MoE, which requires neural networks with significantly different design and performance. Therefore, the focus is on selecting neural networks with different structures and principles, which is why we have chosen CenterNet and Expert2.

\subsection{Performance Evaluation with Two Identical Observation Scenarios}\label{3.5}
To comprehensively demonstrate the efficacy of our algorithm, we have conducted performance evaluations using two identical observational scenarios. The first scenario involves observations conducted against different sky backgrounds, aiming to illustrate our framework's adaptive capability in processing images with various observational qualities. The second scenario utilizes the intersection mode, where the relative motion speed of celestial objects differs across different images. This scenario is designed to showcase our algorithm's versatility in adaptively processing images acquired under varying observational modes. These two scenarios collectively provide a robust test of our algorithm's performance across diverse astronomical imaging conditions. The following subsections will delve into the details of these evaluations, presenting a thorough analysis of our algorithm's performance in each scenario.\\

\subsubsection{Performance Evaluation with Images of Different Sky Background}\label{3.5.1}
\begin{figure}
    \centering
    \includegraphics[scale=0.4]{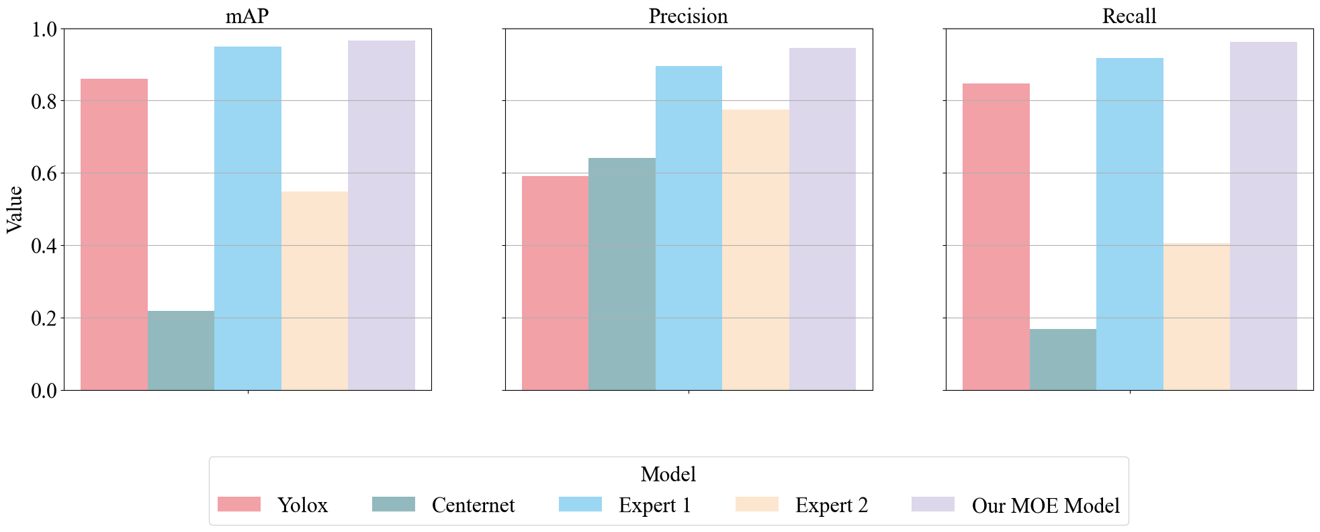}
    \caption{mAP, Precision, and Recall for each model for Test A}
    \label{Figure3.5.1-1}
\end{figure}

\begin{figure}
    \centering
    \includegraphics[scale=0.5]{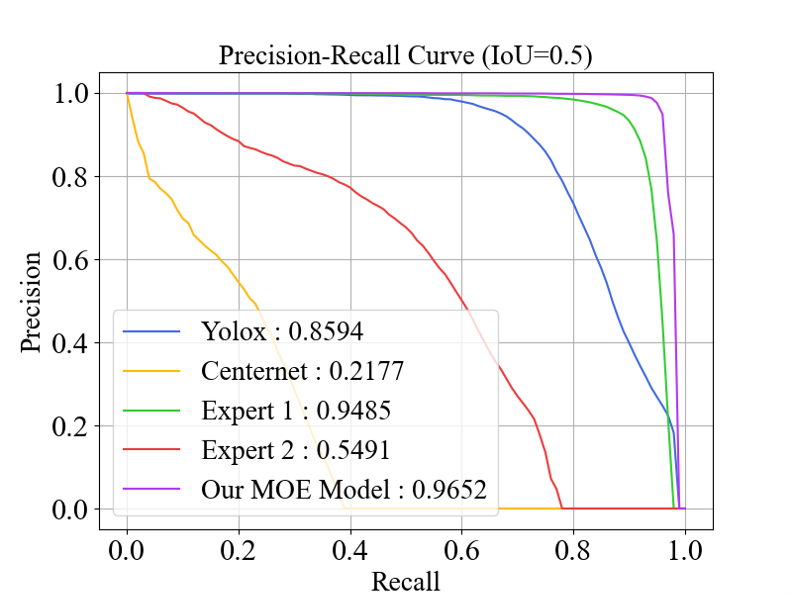}
    \caption{Precision-Recall (PR) Curve for different detection algorithms in Test A with IoU of 0.5.}
    \label{figure3.5.1-2}
\end{figure}

In this subsection, we evaluate the algorithm's performance on a set of images with varying sky background, and we refer to this test as Test A. We set the value range of sky background from 14 to 19 and generate 1500 simulated images under different observation conditions. We use our trained MoE model to detect FMCB within these observation images. For comparative analysis, we also employ E1, E2, YOLOX, and CenterNet for source detection under the same conditions. Table~\ref{table3.5.1} presents the comparative results of these detection methods. As clearly shown in the table, YOLOX and CenterNet exhibit relatively lower performance. YOLOX achieves a precision rate of 59.17\%, a recall rate of 84.71\%, and an mAP of 85.94\%. In contrast, E1, benefiting from the integration of physical-inspired information, achieves a precision rate of 89.45\%, a recall rate of 91.78\%, and an mAP of 94.85\%, outperforming YOLOX in all metrics. Similarly, CenterNet achieves a precision rate of 64.07\%, a recall rate of 16.81\%, and an mAP of 21.77\%, while E2 achieves a precision rate of 77.44\%, a recall rate of 40.47\%, and an mAP of 54.91\%, showing significant improvements over CenterNet in all metrics. Meanwhile, the MoE model outperforms all other models, demonstrating exceptional performance with a precision rate of 94.54\%, a recall rate of 96.13\%, and an mAP of 96.52\%. Figure~\ref{Figure3.5.1-1} and Figure~\ref{figure3.5.1-2} provides a visual representation of these results, clearly illustrating our algorithm's performance. This comprehensive evaluation underscores the robustness and adaptability of our MoE approach in handling varying observational conditions, outperforming both its constituent parts and state-of-the-art alternatives.\\
\begin{table}[ht]
    \centering
    \caption{Comparative analysis of the performance of different detection algorithms in processing Test A}
    \label{table3.5.1}
    \begin{tabular}{|l|c|c|c|}
        \hline
        \textbf{Method} & \textbf{mAP} & \textbf{Precision} & \textbf{Recall} \\
        \hline
        YOLOX & 85.94\% & 59.17\% & 84.71\% \\
        CenterNet & 21.77\% & 64.07\% & 16.81\% \\
        Expert1 & 94.85\% & 89.45\% & 91.78\% \\
        Expert2 & 54.91\% & 77.44\% & 40.47\% \\
        Our MoE Model & 96.52\% & 94.54\% & 96.13\% \\
        \hline
    \end{tabular}
\end{table}

\subsubsection{Performance Evaluation with Images Obtained by the Intersection Observation Mode}\label{3.5.2}
\begin{figure}
    \centering
    \includegraphics[scale=0.4]{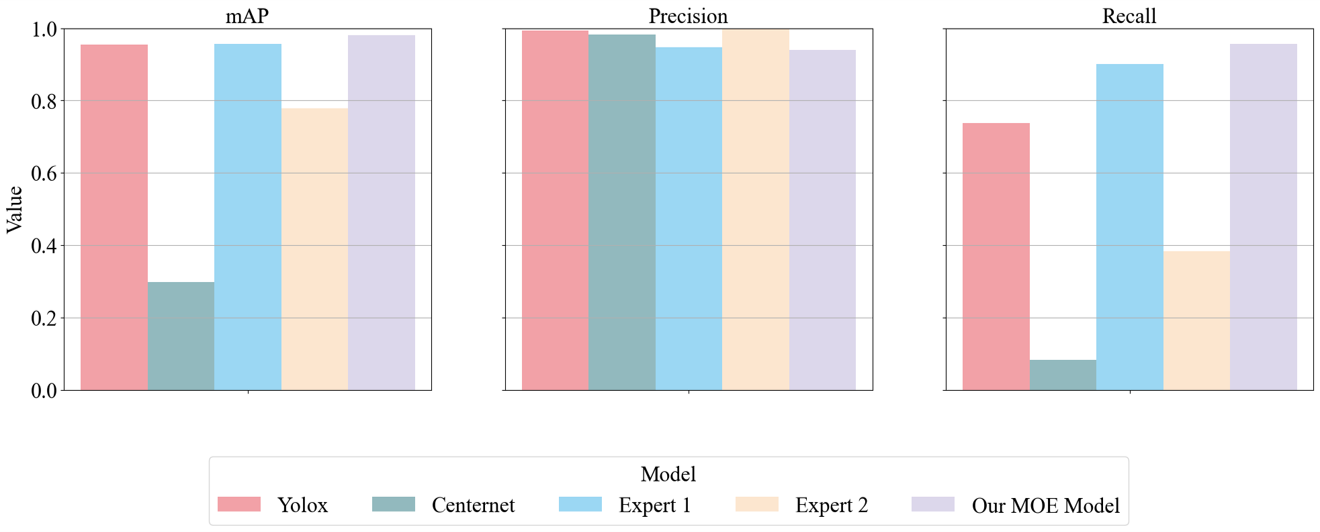}
    \caption{mAP, Precision, and Recall for each model for Test B}
    \label{Figure3.5.2-1}
\end{figure}

\begin{figure}
    \centering
    \includegraphics[scale=0.5]{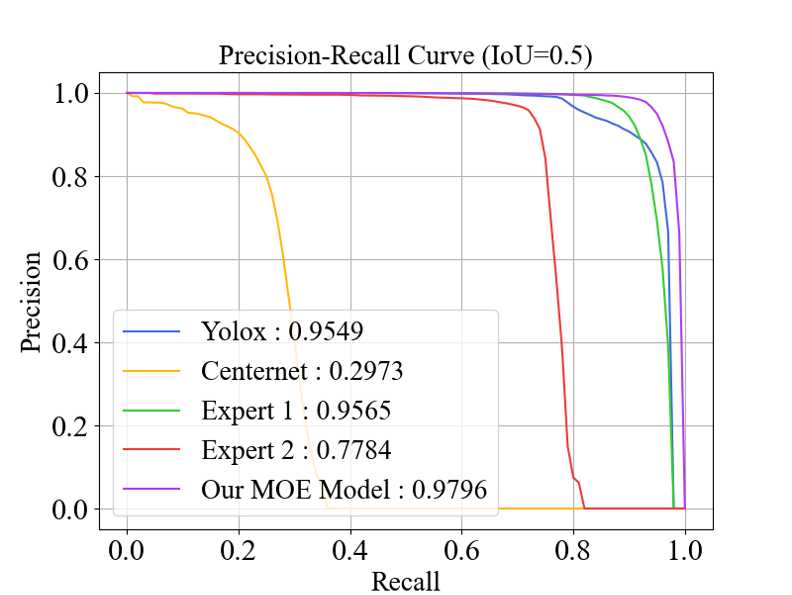}
    \caption{Precision-Recall (PR) Curve for different detection algorithms in Test B with IoU of 0.5}
    \label{figure3.5.2-2}
\end{figure}

\begin{figure}
\centering
\begin{minipage}[b]{0.3\textwidth}
    \centering
    \includegraphics[width=\textwidth]{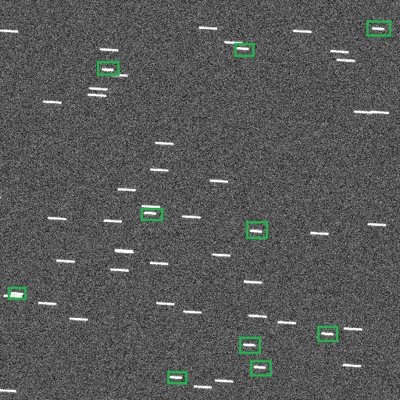}
\end{minipage}
\begin{minipage}[b]{0.3\textwidth}
    \centering
    \includegraphics[width=\textwidth]{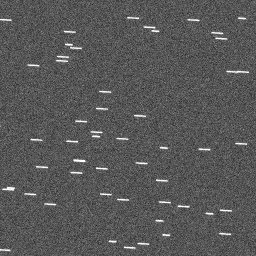}
\end{minipage}\\
\begin{minipage}[b]{0.3\textwidth}
    \centering
    \includegraphics[width=\textwidth]{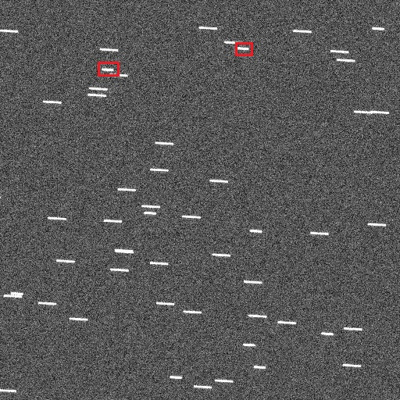}
\end{minipage}
\begin{minipage}[b]{0.3\textwidth}
    \centering
    \includegraphics[width=\textwidth]{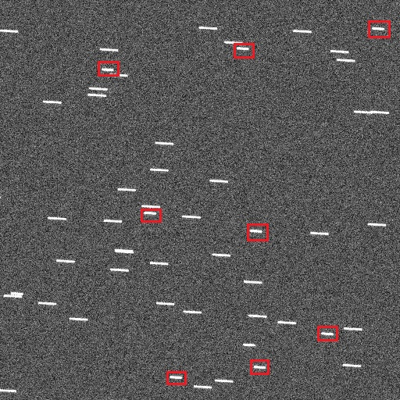}
\end{minipage}
\caption{The green box is the FMCB, and the detection results are represented by the red box. From left to right, top to bottom, this figure shows the ground truth value, the detection results of CenterNet, Yolox, and the method in this paper. As shown in this figure, the method proposed in this paper could effectively detect FMCB.}
\label{Figure3.5.2}
\end{figure}

In this subsection, we employ a more challenging scenario to rigorously test our algorithm's performance, and we refer to this test as Test B. The intersection mode presents an exceptionally demanding environment for FMCB detection algorithms due to the extreme variations in the shapes of background stars and FMCB between different images. To simulate this complexity, we have generated simulated images with randomly distributed moving directions from 0 deg to 180 deg and relative speeds ranging from 1.53 to 3.38 pixels per frame. We then apply our trained MoE model, along with E1, E2, YOLOX, and CenterNet, directly to these simulated observation images for FMCB detection. Table~\ref{table3.5.2} presents the comparative results of these detection methods. It can be observed that Expert1, which integrates physical-inspired information, achieves overall better performance than YOLOX. Similarly, Expert 2 demonstrates superior overall performance compared to CenterNet. Our approach demonstrates remarkable efficacy, achieving a precision rate of 93.96\%, a recall rate of 95.56\%, and a mAP of 97.96\%. These results underscore the significant performance enhancement provided by our physical-inspired information integration. Figure~\ref{Figure3.5.2-1} and Figure~\ref{figure3.5.2-2} showcase the performance of our algorithm. Figure~\ref{Figure3.5.2} showcases several challenging simulated images where classical methods failed to achieve effective detection. In contrast, our trained neural network consistently delivers better results in detecting FMCB, regardless of variations in observational modes. This robust performance across diverse scenarios highlights the adaptability and effectiveness of our approach in addressing the complex challenges posed by astronomical imaging in the intersection mode.\\

\begin{table}[ht]
    \centering
    \caption{Comparative analysis of the performance of different detection algorithms in processing Test B}
    \label{table3.5.2}
    \begin{tabular}{|l|c|c|c|}
        \hline
        \textbf{Method} & \textbf{mAP} & \textbf{Precision} & \textbf{Recall} \\
        \hline
        YOLOX & 95.49\% & 99.34\% & 73.72\% \\
        CenterNet & 29.73\% & 98.18\% & 8.17\% \\
        Expert1 & 95.65\% & 94.64\% & 90.09\% \\
        Expert2 & 77.84\% & 99.74\% & 38.25\% \\
        Our MoE Model & 97.96\% & 93.96\% & 95.56\% \\
        \hline
    \end{tabular}
\end{table}

\subsection{Performance Evaluation with Real Observation Data}\label{3.6}
\begin{figure*}
    \centering
    \begin{minipage}{0.3\textwidth}
        \centering
        \includegraphics[width=\linewidth]{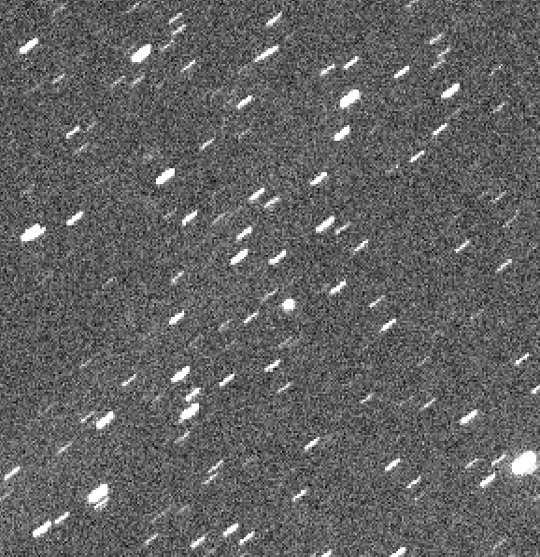}
        \centering\footnotesize\textbf{(a)} 
    \end{minipage}%
    \hspace{0.05\textwidth}%
    \begin{minipage}{0.3\textwidth}
        \centering
        \includegraphics[width=\linewidth]{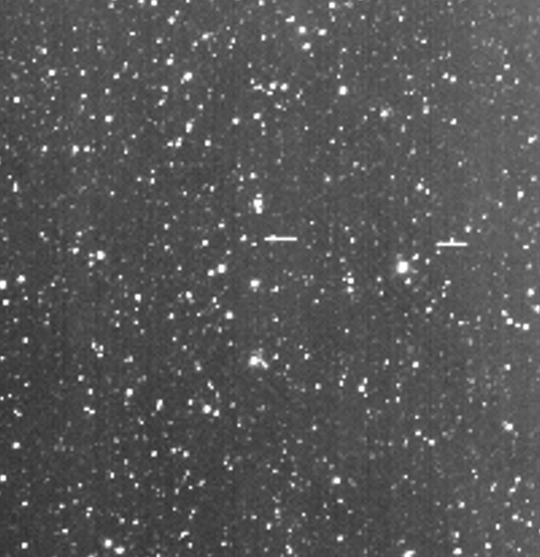}
        \centering\footnotesize\textbf{(b)}
    \end{minipage}
    \caption{Figure(a) shows the real data under Tracking Mode, while Figure(b) shows the real data under Sidereal Mode.} 
    \label{figure:realdata} 
\end{figure*}

\begin{figure}[ht]
    \centering
    \includegraphics[scale=0.5]{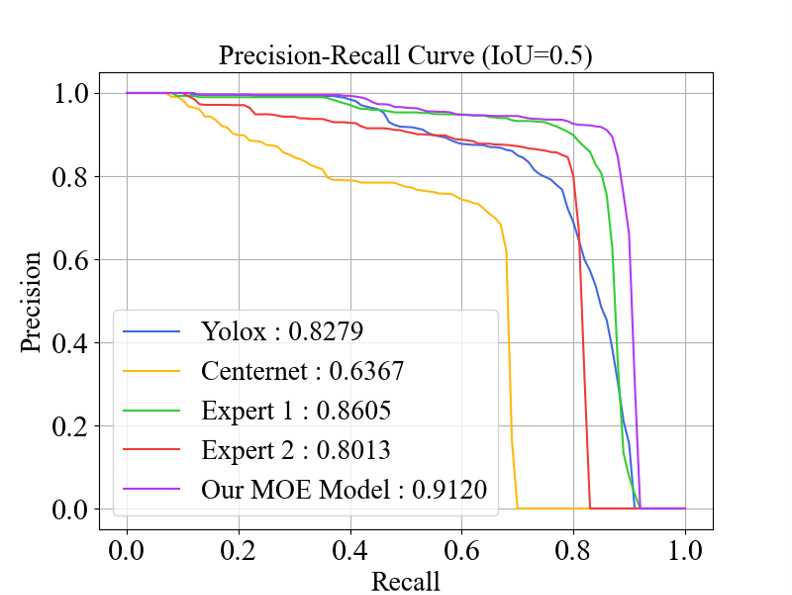}
    \caption{Precision-Recall (PR) Curve for different detection algorithms in the real data with IoU of 0.5.}
    \label{figure3.6.1}
\end{figure}

At last, we have collected a set of real observation data obtained by two different telescopes with different observation modes operated by the Purple Mountain Observatory, as shown in Figure \ref{figure:realdata}. The dataset includes 2400 images obtained by Tracking Mode and Sidereal Mode, and these images are in a 1:1 ratio. We have split these images into fine-tuning, validation, and test sets with a ratio of 7:3:3 and set the number of epochs for fine-tuning to 100. Table \ref{table3.6.1} shows the detection performance of different models on the overall dataset. It should be noted that all these models are trained and tested with aforementioned dataset and we have tested their performance after they have achieved their best performance. As shown in the table, our MoE model achieves the highest mAP (91.20\%) and Recall (90.46\%), and also performs well in Precision (90.96\%). This indicates that the MoE model performs strongly across all metrics. Figure~\ref{figure3.6.1} shows the PR curves of various models, which also shows that the MoE model has the best performance, because the AP is 0.9120, which is the largest value among all these models.\\

\begin{table}
\centering
\begin{tabular}{|l|c|c|c|}
\hline
\textbf{Method} & \textbf{mAP} & \textbf{Precision} & \textbf{Recall} \\
\hline
YOLOX         & 82.79\%      & 85.09\%           & 76.97\%         \\
CenterNet     & 63.67\%      & 80.44\%           & 60.37\%         \\
Expert1       & 86.05\%      & 88.93\%           & 83.30\%         \\
Expert2       & 80.13\%      & 91.00\%           & 68.62\%         \\
Our MoE Model & \textbf{91.20\%} & \textbf{90.96\%} & \textbf{90.46\%} \\
\hline
\end{tabular}
\caption{Detection results of various models on real observation data}
\label{table3.6.1}
\end{table}

The performance of these models in detecting FMCB across various observation modes has been further examined. The detection results of each model under different observation modes are presented in Table~\ref{table3.6.2} and Figure~\ref{figure3.6.1}. It is evident that the Sidereal Mode and Tracking Mode significantly influence the models' effectiveness, with certain models, such as YOLOX and CenterNet, demonstrating markedly superior performance in one mode over the other. Notably, our MoE model consistently excels in both Sidereal and Tracking Modes, particularly in the Tracking Mode, where it surpasses all other models. These findings suggest that our MoE model displays strong adaptability and high detection accuracy across various observation conditions, rendering it suitable for practical astronomical observation tasks. Thus, it is clear that our method achieves the highest performance in detecting real data. 

\begin{table}[ht]
    \centering
    \caption{Detection results of various models in processing images with different observation modes}
    \label{table3.6.2}
    \begin{tabular}{|l|c|c|c|c|c|c|}
        \hline
        \textbf{Method} & \multicolumn{3}{c|}{\textbf{Sidereal Mode}} & \multicolumn{3}{c|}{\textbf{Tracking Mode}} \\
        \cline{2-7}
                        & \textbf{mAP} & \textbf{Precision} & \textbf{Recall} & \textbf{mAP} & \textbf{Precision} & \textbf{Recall} \\
        \hline
        YOLOX           & 87.50\%      & 96.26\%           & 82.77\%         & 72.93\%      & 75.56\%           & 71.53\% \\
        CenterNet       & 85.62\%      & 97.33\%           & 75.95\%         & 39.58\%      & 63.30\%           & 45.73\% \\
        Expert1         & 85.23\%      & 91.70\%           & 83.71\%         & 86.27\%      & 86.46\%           & 82.92\% \\
        Expert2         & 85.97\%      & 97.04\%           & 74.62\%         & 72.55\%      & 85.10\%           & 62.99\% \\
        Our MoE Model   & \textbf{90.66\%} & \textbf{93.41\%} & \textbf{88.64\%} & \textbf{91.72\%} & \textbf{88.85\%} & \textbf{92.17\%} \\
        \hline
    \end{tabular}
\end{table}

\section{Conclusion and Future Work} \label{sec4}
FMCB are important for studying of celestial objects within the solar system. Detection of these celestial objects is the first and most important step. As more and more different observations are carried out by space and ground based telescopes, more and more observation images are obtained by different observation modes. Although deep learning based algorithms are promising in detection of celestial objects from ordinary observation images, its performance will drop down, when we consider different observation conditions and modes. In this paper, we propose a novel method, which integrates physical-inspired information as prior information, for source detection. Our method could adaptively process images obtained by different observation modes and detect celestial objects that have different moving modes than back ground stars. To further increase the performance of our algorithm, we propose to use the MoE to integrate two PiNNs for FMCB detection. Results show that our method is effective and has better performance than other methods.\\

While our method demonstrates effectiveness in detecting FMCB across various observational modes, several challenges remain to be addressed. The observational scenarios addressed by our current methodology remain somewhat constrained. In practical applications, it is essential to account for cosmic rays and other forms of noise that are characteristic of all astronomical observations. Under these conditions, there is a significant likelihood of generating false detection results. Consequently, our algorithm may serve as an initial step for detection; thereafter, we could utilize adjacent images to track rapidly moving celestial objects, thereby enhancing detection efficiency. This aspect will be elaborated upon in our forthcoming papers. Furthermore, our current implementation of the MoE model incorporates only two neural networks. Given that the MoE architecture can theoretically integrate an infinite number of experts, we aim to expand our approach by incorporating additional FMCB detection neural networks in future iterations of our source detection algorithm. This expansion has the potential to further enhance the versatility and accuracy of our detection system across a broader range of astronomical imaging scenarios. Additionally, for object detection tasks with real data, challenges such as uneven data quality, large data volumes, high labeling costs, class imbalance, and tool limitations pose significant difficulties. To address these, in the future, we can combine a citizen science labeling platform with semi-supervised learning. Initially, a small high-quality labeled dataset from the platform trains a model, which then generates pseudo-labels for unlabeled data. Volunteers on the platform verify and correct these pseudo-labels, and the refined data retrains the model. Iterating this process establishes an efficient real data labeling pipeline. By addressing these challenges, we anticipate significant improvements in the robustness and applicability of our method for detecting FMCB in diverse observational contexts.\\

\section*{Acknowledgments} \label{acknowledgments}
The code used in this paper is shared in the PaperData Repository powered by the China-VO with the following \href{https://nadc.china-vo.org/res/r101585/}{link}. This work is supported by the National Key R \& D Program of China (No. 2023YFF0725300) and the National Natural Science Foundation of China (NSFC) with funding numbers 12173027. This work is supported by the Young Data Scientist Project of the National Astronomical Data Center.\\


%

\vspace{5mm}


\software{SkyMaker \citep{bertin2009skymaker},
          Source Extractor \citep{bertin1996sextractor},
          astropy \citep{robitaille2013astropy},
          Matplotlib \citep{Hunter2007},
          pytorch \citep{paszke2019pytorch},
          PNet \citep{sun2023pnet}
          }



\bibliography{sample631}{}
\bibliographystyle{aasjournal}



\end{document}